\documentclass[sigplan,10pt]{acmart}
\acmSubmissionID{491}
\acmConference[EuroSys '25]{EuroSys 2025}{March 30--April 03, 2025}{Rotterdam, NL}

\usepackage[]{hyperref}
\usepackage{algorithm}
\usepackage{algpseudocode}
\usepackage{caption}
\usepackage{subcaption}
\usepackage{multirow}
\usepackage{array}
\usepackage{float}
\usepackage{stfloats}
\usepackage{tabularx}
\usepackage{tabulary}
\usepackage[normalem]{ulem}
\usepackage{subfiles}
\usepackage{siunitx}
\usepackage{enumitem}
\RequirePackage[loading]{tracefnt}

\algrenewcommand\algorithmicindent{1em}%
\newcolumntype{Y}{>{\centering\arraybackslash}X}

\setcopyright{rightsretained}

\begin{document}

\acmYear{2025}\copyrightyear{2025}
\acmConference[EuroSys '25]{Twentieth European Conference on Computer Systems}{March 30--April 3, 2025}{Rotterdam, Netherlands}
\acmBooktitle{Twentieth European Conference on Computer Systems (EuroSys '25), March 30--April 3, 2025, Rotterdam, Netherlands}
\acmDOI{10.1145/3689031.3717483}
\acmISBN{979-8-4007-1196-1/25/03}

\ccsdesc[500]{Computer systems organization~Cloud computing}
\ccsdesc[500]{Applied computing~Data centers}

\keywords{Cloud Computing, Cluster Scheduling}

\title{Eva: Cost-Efficient Cloud-Based Cluster Scheduling}

\author{Tzu-Tao Chang}
\email{tchang85@wisc.edu}
\affiliation{%
  \institution{University of Wisconsin-Madison}
  \country{USA}
}
\author{Shivaram Venkataraman}
\email{shivaram@cs.wisc.edu}
\affiliation{%
  \institution{University of Wisconsin-Madison}
  \country{USA}
}

\begin{abstract}
Cloud computing offers flexibility in resource provisioning, allowing an organization to host its batch processing workloads cost-efficiently by dynamically scaling the size and composition of a cloud-based cluster -- a collection of instances provisioned from the cloud. However, existing schedulers fail to minimize total cost due to suboptimal task and instance scheduling strategies, interference between co-located tasks, and instance provisioning overheads. We present Eva, a scheduler for cloud-based clusters that reduces the overall cost of hosting long-running batch jobs. Eva leverages reservation price from economics to derive the optimal set of instances to provision and task-to-instance assignments. Eva also takes into account performance degradation when co-locating tasks and quantitatively evaluates the trade-off between short-term migration overhead and long-term provision savings when considering a change in cluster configuration. Experiments on AWS EC2 and large-scale trace-driven simulations demonstrate that Eva reduces costs by 42\% while incurring only a 15\% increase in JCT, compared to provisioning a separate instance for each task.
\end{abstract}

\maketitle 

\section{Introduction}
\label{sec:intro}
Cloud computing has seen widespread adoption, with demand continually increasing due to the rise of emerging technologies such as machine learning (ML) and big data analytics~\cite{forbes2023cloudgrowth, google2023cloudtrends}. Specifically, for batch computing workloads, the flexibility and scalability of cloud platforms offers a solution for organizations to host jobs in a cost-efficient manner~\cite{dageville2016snowflake} using a \textit{cloud-based cluster}, i.e., a pool of instances provisioned from the cloud. 
As a result, research institutions and enterprises have migrated their batch processing workloads from internal computing clusters to cloud-based clusters consisting of hundreds or thousands of cloud instances~\cite{aws2020adelaide, aws2021discovery, google2021arabesqueAI, microsoft2023bath}.

To ensure cost-efficiency with a cloud-based cluster, effective scheduling mechanisms are necessary to map submitted tasks to appropriate instances~\cite{alipourfard2017cherrypick}. While task scheduling for batch processing workloads has been extensively studied in the fixed-sized cluster setup~\cite{verma2015borg, hindman2011mesos, vavilapalli2013yarn, grandl2014tetris, gu2019tiresias, mahajan2020themis, li2023lyra, narayanan2020gavel, xiao2018gandiva, mohan2022synergy, qiao2021pollux}, the additional flexibility in cloud-based clusters introduces complexity to the scheduling problem. Specifically, on-demand provisioning of resources removes the time jobs spend waiting in the queue due to insufficient resources~\cite{chung2018stratus}, which is the primary focus of the majority of fixed-sized cluster schedulers. In addition, a cloud-based cluster can dynamically adjust its composition by leveraging the diverse range of heterogeneous instances offered by the cloud provider, with each instance type having its own cost. As a result, these factors change the objective of the scheduling problem from only minimizing job completion time (JCT) to minimizing total provisioning cost without compromising job throughput. Since task scheduling and instance provisioning are fundamentally linked -- tasks should be scheduled to efficiently utilize the resources available from the provisioned instances, whereas the instances should be selected to match the demands of the tasks -- the two aspects should be jointly optimized to determine the optimal \textit{cluster configuration}, which includes the quantity and types of instances that compose the cluster and the task-to-instance assignment.

\begin{figure}
    \centering
    \includegraphics[width=0.7\linewidth]{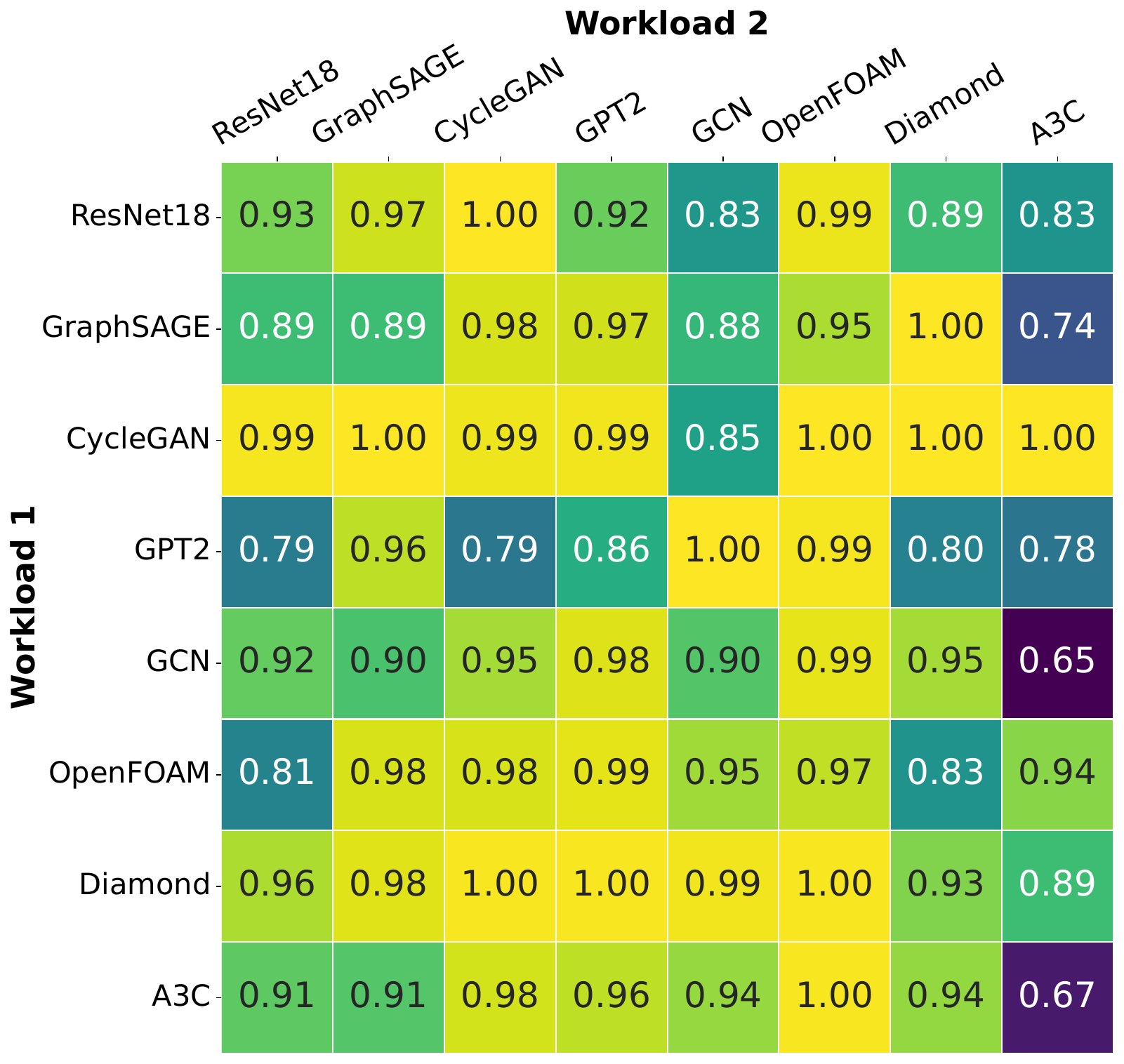}
    \caption{Performance of batch jobs when co-located on the same instance. Each cell shows the normalized throughput of Workload 1 when co-located with Workload 2. Both workloads receive the resources they requested, as listed in Table~\ref{tab:evaluate-workload}, and are assigned to separate GPUs and CPUs on the same instance. The jobs start simultaneously and run for 10 minutes. Throughput is measured for each job during this period and normalized by dividing it by the job's standalone throughput on an instance without co-location.}
    \label{fig:tput-matrix}
\end{figure}

In light of this, prior work has proposed schedulers for the cloud setting~\cite{chung2018stratus, shastri2017hotspot, harlap2017proteus, yang2023skypilot}. However, they fail to address certain challenges that are essential for cost-efficient hosting of batch jobs (\S\ref{subsec:background/use_case}).
First, resource demands are diverse and weakly correlated across batch processing jobs in a cluster~\cite{grandl2014tetris, zhang2014harmony}, which provides opportunity to co-locate multiple tasks onto the same instance to reduce the number of instances provisioned and thus lowering total cost. However, interference between co-located tasks results in performance degradation, which can vary significantly across different tasks. Figure~\ref{fig:tput-matrix} shows that performance degradation can range from 0-36\% for just two co-located tasks. As a result, naïvely co-locating tasks can lead to significantly longer job durations, which in turn increases instance uptime and results in higher overall provisioning costs. In addition, the optimal cluster configuration can change over time as jobs are submitted to or complete in the system. \textit{Cluster reconfigurations}, i.e., switching from one cluster configuration to another, can lead to more cost-efficient resource provisioning but involves task migrations and instance launches, which introduce delays on the order of minutes, as shown in Table~\ref{tab:delay}. During these delays, provisioned resources remain idle, leading to wasted cost. Consequently, the scheduler must consider the trade-off between long-term provisioning cost saving and short-term migration overhead.

To address these challenges, we present Eva, a cluster scheduler that aims at serving batch computing workloads cost-efficiently in a cloud-based cluster. In Eva, we propose packing tasks into a set of instances to improve utilization and reduce cost. To link task scheduling and instance provisioning, Eva's scheduling algorithms draw insight from an effective heuristic to the variable sized bin packing problem (VSBPP), which is known to be NP-hard~\cite{friesen1986vsbpp}. The heuristic prioritizes larger bin types and balls to reduce the number of used bins and unused space within each bin, lowering the total cost. While effective in the single-dimensional setting, generalizing the heuristic for cloud-based cluster scheduling introduces challenges due to the presence of multi-dimensional resources (e.g., GPU, CPU, RAM), making it difficult to define a single ``size'' for instance types and tasks. 

In Eva, we capture the intuition of minimizing resource fragmentation through \emph{cost}, which is proportional to the quantity and type of resources involved (\S\ref{subsec:scheduling/full-reconfig}). Specifically, tasks are considered in descending \textit{reservation price}~\cite{steedman2008reservationprice}, a concept borrowed from economics that represents the maximum price a buyer is willing to pay for a good or a service, while instance types are considered in descending hourly cost. In the context of scheduling, the reservation price of executing a task is the hourly cost that would be incurred if the task were executed on a standalone instance without co-location. This provides a basis for evaluating the cost-efficiency of a task-to-instance assignment: the sum of the reservation prices of a set of tasks assigned to an instance should be no less than the actual hourly cost of the instance.

To account for performance degradation caused by co-location interference, we extend reservation price to consider the throughput of a task (\S\ref{subsec:scheduling/interference}). The \textit{throughput-normalized reservation price} of a task represents the maximum price the user is willing to pay to host the task at  a certain throughput level under interference. For example, if a task can be hosted on an instance type that costs \$3 per hour and achieves 100\% throughput without co-location, the user would be willing to pay \(\$3 \times 0.8 = \$2.40\) per hour when its throughput decreases to 80\% due to interference from co-locating with other tasks. This allows us to perform the same cost-efficiency evaluation of a task-to-instance assignment in terms of performance, even in the presence of multi-task jobs, where the performance of tasks within the same job are interdependent  (\S\ref{subsec:scheduling/multi-task}).

Based on throughput-normalized reservation price, we design two scheduling algorithms: Full Reconfiguration (\S\ref{subsec:scheduling/full-reconfig}) and Partial Reconfiguration (\S\ref{subsec:scheduling/migration}). These algorithms are used in combination to update the cluster configuration online. Full Reconfiguration considers all tasks currently in the system to determine the cluster configuration that leads to minimal provisioning cost. In contrast, Partial Reconfiguration preserves the majority of the current cluster configuration and updates only a subset of tasks and instances to minimize migration overhead. At each scheduling round, Eva runs both algorithms to generate two candidate cluster configurations, from which Eva selects one. Intuitively, Full Reconfiguration is preferred when the potential cost savings in provisioning justify the incurred migration overhead, particularly if these savings are substantial and long-lasting. We propose a quantitative method (\S\ref{subsec:scheduling/migration}) to estimate the trade-off between provisioning cost saving and migration overhead, which Eva uses as the criterion to choose between the two candidate cluster configurations.

We have implemented Eva along with a high-fidelity simulator in Python. While the current implementation assumes AWS EC2~\cite{aws-ec2} as the cloud platform, Eva's modular design (\S\ref{sec:design}) ensures easy adaptation for other cloud providers. Tasks are executed as containers in the cloud, ensuring no limitations on frameworks or task environments. Additionally, Eva includes a lightweight iterator API to monitor job throughput, requiring minimal code changes on the user side. 

We evaluate Eva on AWS EC2 with a trace spanning various batch applications in ML and scientific computing (Table~\ref{tab:evaluate-workload}). We find that Eva reduces total cost by 25\% and increases average resource allocation by 1.2\(\times\). Further, our simulations using Alibaba production trace~\cite{weng2023trace} of more than 6,200 jobs show that Eva reduces cost by 42\% and consistently achieves significant cost reductions even in adverse scenarios with high co-location interference and task migration delays.
\vspace{-2ex}
\section{Background and Motivation}
\label{sec:background}
\subsection{Scheduling Batch Processing Workloads}
\label{subsec:background/problem_setting}
Batch processing workloads, such as ML training, are increasingly prevalent in both research and enterprise production environments~\cite{gu2019tiresias, jeon2019philly}. These workloads are resource-intensive and can run for extended periods, ranging from hours to days~\cite{jeon2019philly, xiao2018gandiva}. Organizations have hosted these workloads on dedicated, fixed-sized clusters~\cite{gu2019tiresias, mohan2022synergy, xiao2018gandiva, xiao2020antman, jeon2019philly}, which are managed by schedulers to optimize resource allocation and job scheduling. Traditional schedulers such as Mesos~\cite{hindman2011mesos}, YARN~\cite{vavilapalli2013yarn}, Tetris~\cite{grandl2014tetris}, and Borg~\cite{verma2015borg} are used to serve CPU-intensive big data workloads such as MapReduce~\cite{dean2004mapreduce} jobs. To meet the increasing popularity and demand for ML training, numerous cluster schedulers tailored to the unique characteristics and constraints of ML jobs have been proposed~\cite{gu2019tiresias, mahajan2020themis, li2023lyra, narayanan2020gavel, xiao2018gandiva, mohan2022synergy, qiao2021pollux}. These schedulers focus on efficiently utilizing costly accelerators such as GPUs, but share the same overarching goal: scheduling jobs to minimize JCT and maximize resource utilization. Prior research has showed that the aggregate resource demands of big data and ML applications within a cluster are bursty and fluctuate over time~\cite{mahajan2020themis, grandl2014tetris}, leading to under-utilization and inefficient usage of expensive resources in a fixed-sized cluster setup. 

\subsection{Batch Computing in the Cloud}
\label{subsec:background/batch_computing}
With the flexibility to dynamically scale and adjust computing resources on-demand, cloud computing has been widely adopted and continues to grow~\cite{gartner2023cloudspending}. The pay-as-you-go pricing model offers opportunities for cost-efficient hosting of emerging batch processing workloads using a cloud-based cluster. As a result, organizations have begun migrating batch job computations to the cloud~\cite{weng2022mlaas, sagemaker}. For example, research institutions have leveraged cloud-based clusters with over 2,000 instances for bioinformatics computations~\cite{aws2020adelaide}, while enterprises have moved their ML training workloads to the cloud to reduce costs~\cite{google2021arabesqueAI}.

However, the additional flexibility to provision resources from a pool of heterogeneous instances in a cloud-based cluster introduces complexity and alters the job scheduling problem. With the ability to scale and change the resource composition of the cluster, the focus shifts away from only minimizing JCT to minimizing total provisioning cost without compromising job throughput. This motivates us to design an effective scheduler for cloud-based clusters.

\subsection{Target Use Case and Problem Formulation}
\label{subsec:background/use_case}
\textit{Consider an enterprise with multiple ML development teams, each regularly training ML models on the cloud. Initially, each team creates appropriate instances for their specific jobs. Since the enterprise's total cloud costs depend on the duration and type of instances provisioned, the enterprise seeks to minimize overall expenses. Thus, the enterprise decides to create a shared cloud-based cluster where teams can submit their ML training jobs for execution.}

\textit{With a shared cloud-based cluster, the enterprise aims to select appropriate instances and efficiently assign jobs from different teams. Since all teams belong to the same enterprise, security concerns with instance sharing are not an issue.}

Formally, we consider the following problem: in a cloud-based cluster, users submit batch jobs that consist of one or more tasks to be executed on cloud instances. Let $\mathcal{T}$ denote the set of all tasks, where a task $\tau \in \mathcal{T}$ has a demand $D_\tau^r$ for resource $r$. Given a set of available instance types $\mathcal{K}$ with no limit on the number of instances to provision from each type, and instance type $k$ has a capacity $Q_k^r$ for resource $r$, along with an hourly cost of $C_k$, the objective is to accommodate $\mathcal{T}$ at minimal cost while maintaining job throughput comparable to that of a dedicated, non-shared cloud-based cluster.

Below, we outline three primary challenges and opportunities that are crucial for cost-efficient hosting of batch processing jobs in the cloud. While there has been prior work that attempted to address the aforementioned problem~\cite{chung2018stratus, shastri2017hotspot, harlap2017proteus, xu2019ispot, yang2023skypilot}, they fall short in tackling these challenges, underscoring the need for a new scheduler.\\
\textbf{Varied Resource Usage} Batch processing jobs exhibit diverse resource demands~\cite{grandl2014tetris}. We profiled and present the resource demands of 10 batch processing jobs from different applications in Table~\ref{tab:evaluate-workload}. Notably, while ML training jobs commonly rely on accelerators like GPUs, the specific GPU requirements vary between models due to factors such as model size and scalability. In addition, ML tasks involving image models benefit from increased CPU capacity for efficient data pre-processing~\cite{mohan2022synergy}, whereas graph learning tasks require substantial amounts of RAM for storing and accessing large embedding tables~\cite{mohoney2021marius}.

In addition, prior work has shown a lack of correlation between resource demands across tasks in production clusters~\cite{grandl2014tetris, zhang2014harmony}. This implies that tasks with complementary resource demands can be \textit{co-located} to reduce the amount of idle resources. Fixed-sized cluster schedulers such as Tetris~\cite{grandl2014tetris} and Synergy~\cite{mohan2022synergy} use scheduling algorithms that leverage task co-location to reduce job queuing delays. In the context of cloud-based clusters, task co-location improves resource allocation and reduces the number of instances needed to be provisioned, thereby lowering total cost. However, this is not considered by most existing cloud managers~\cite{shastri2017hotspot, harlap2017proteus, xu2019ispot, yang2023skypilot}.\\
\textbf{Co-location Interference} While co-locating tasks reduces the number of instances that need to be provisioned, tasks co-located on the same cloud instance inevitably share low-level resources such as last-level cache (LLC), disk I/O bandwidth, and network bandwidth~\cite{cortez2017rc, zhuravlev2010contention,  qiu2020firm}, which cloud users have limited control over. Contention for shared resources leads to interference among co-located jobs, resulting in performance degradation. The degradation could potentially increase job duration and thus instance uptime, leading to higher total cost. As a result, we believe cloud-based cluster schedulers must be interference-aware and take performance into consideration when making scheduling decisions.

Prior work has attempted to incorporate the effect of co-location interference in cluster schedulers. Paragon~\cite{delimitrou2013paragon} and Quasar~\cite{delimitrou2014quasar} estimate the impact of interference on a task using collaborative filtering, while Owl~\cite{tian2022owl} directly profiles the interference beforehand. Based on these information, their schedulers avoid co-locating tasks that could cause severe interference with each other, thereby meeting user-specified quality of service. Since provisioning costs are not directly accounted for in the scheduling algorithms, the resulting configuration might not be cost-efficient in terms of dollar-normalized throughput. Eva aims to address this by linking costs with performance when co-locating tasks. \\
\begin{table}[tbp]
    \centering
    \begin{tabular}{l|cc}
        \hline
        Delay Type & Delay (sec) & Average (sec) \\ \hline
        Instance Acquisition & 6 -- 83 & 19 \\ 
        Instance Setup & 140 -- 251 & 190\\
        Job Checkpointing & 2 -- 30 & 8 \\
        Job Launching & 1 -- 160 & 47 \\ \hline
    \end{tabular}
    \caption{Reconfiguration Delays. Instance-related delays are based on measurements from 126 instances on AWS EC2, while the job-related delays are measured from 120 jobs sampled across the 10 workloads listed in Table~\ref{tab:evaluate-workload}.}
    \label{tab:delay}
    \vspace{-3ex}
\end{table}
\textbf{Quantitative Criterion for Cluster Reconfiguration} As jobs are submitted to or complete in the system, the optimal cluster configuration that minimizes provisioning cost can change over time, so performing cluster reconfigurations can lower provisioning cost. These reconfigurations can introduce non-negligible migration overhead. As the task-to-instance assignment changes, tasks have to be stopped and checkpointed on the original instance and then restarted on another instance. Table~\ref{tab:delay} shows the delays we observed during task migration, which can take up to several minutes. These delays leave resources provisioned but idle, resulting in extra cost. Consequently, existing cloud scheduler such as Stratus~\cite{chung2018stratus} tend to avoid task migration as much as possible. However, the reduction in instantaneous provisioning costs could accumulate to significant cost savings over an extended period, especially for long-running batch workloads. In such cases, a conservative migration strategy is suboptimal. It is thus important to have a quantitative approach to assess the trade-offs between migration overhead and provisioning cost savings with regard to cluster reconfiguration for minimizing total cost.

In summary, we aim to design a scheduler that jointly optimizes task scheduling and instance provisioning to achieve high cost-efficiency for cloud-based clusters.
\vspace{-2ex}
\section{Design}
\label{sec:design}
\begin{figure}
    \centering
    \includegraphics[width=\linewidth]{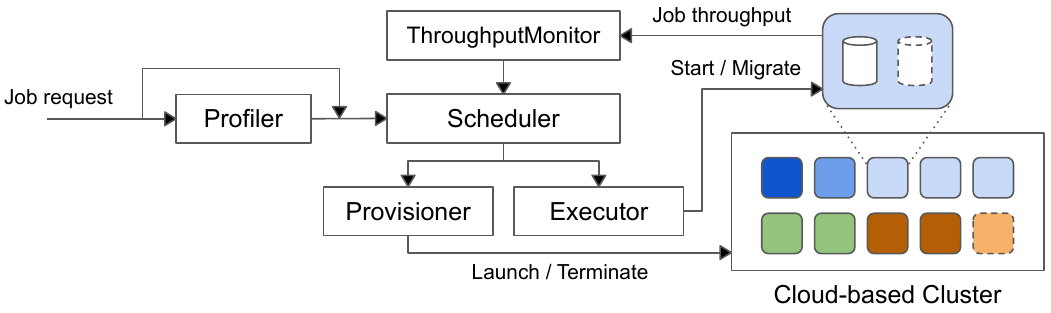}
    \caption{Eva architecture.}
    \label{fig:eva-arch}
\end{figure}
Eva is a cluster scheduler that enables cloud users to cost-efficiently host their batch jobs on a cloud-based cluster. Figure~\ref{fig:eva-arch} illustrates Eva's architecture. A job submitted to Eva consists of one or more tasks that need to be executed on cloud instances. It has resource demands per task for GPU, CPU, and RAM, and, optionally provides the throughput achieved when each task runs on a standalone instance without co-location. The throughput can be estimated using the \texttt{Profiler} if not provided.

Eva performs periodic scheduling. At the end of each scheduling period (e.g. 5 minutes), the \texttt{Scheduler} determines the cluster configuration, including the number of instances, the type of each instance, and the assignment of tasks to instances. Based on the configuration, the \texttt{Provisioner} launches and terminates instances from the cloud provider, while the \texttt{Executor} launches and migrates tasks on these instances. 

Tasks co-located on the same instance are assigned disjoint sets of computing resources (GPU, CPU) but inevitably share lower-level resources like LLC and disk I/O bandwidth, which can lead to interference that degrades performance and reduces cost-efficiency. As a result, the \texttt{Scheduler} needs to know how much a task's throughput is affected when it is co-located with other tasks. Obtaining this information through extensive profiling involves costs that grow exponentially with the number of task types in the system. Instead, the \texttt{ThroughputMonitor} tracks and learn this online, maintaining the co-location throughput table, a data structure recording the throughput of co-located tasks. The table is used by the \texttt{Scheduler} for interference-aware scheduling.

We next elaborate on the \texttt{Scheduler}, the core of Eva that determines the cluster configuration based on resource demands, co-location interference and the trade-off between provisioning cost savings and migration overhead.
\vspace{-1ex}
\section{Scheduling Algorithm}
\label{sec:scheduling}
We first describe an integer linear programming (ILP) formulation of the scheduling problem (\S\ref{subsec:scheduling/ilp}). 
The high computational cost of solving the ILP makes it impractical for real-world deployment. In light of this, Eva employs an effective heuristic scheduling algorithm, which utilizes the concept of reservation price to evaluate the cost-efficiency of task-to-instance assignments (\S\ref{subsec:scheduling/full-reconfig}). To ensure practicality, we explain how to extend the heuristic to account for co-location interference (\S\ref{subsec:scheduling/interference}),  the interdependency between tasks of a multi-task job (\S\ref{subsec:scheduling/multi-task}), and the trade-off between migration overhead and provisioning savings (\S\ref{subsec:scheduling/migration}).
\subsection{ILP Formulation}
\label{subsec:scheduling/ilp}
\begin{table}[tbp]
    \centering
    \label{tab:notation}
    \begin{tabular}{cl}
        \hline
         Symbol & Definition \\
         \hline
         $\mathcal{I}$ & The set of server instances. \\
         $\mathcal{K}$ & The set of available instance types. \\
         $\mathcal{R}$ & The set of resource types.\\
         $\mathcal{T}$ & The set of tasks.\\
         $D^r_\tau$ & The demand for resource $r$ of task $\tau$.\\
         $Q^r_k$ & The capacity of resource $r$ on instance type $k$.\\
         $C_k$ & The cost of instance type $k$.\\
         $x_{ik} \in \{0, 1\}$ & Whether instance $i$ is of the instance type $k$.\\
         $y_{i\tau} \in \{0, 1\}$ & Whether task $\tau$ is assigned to instance $i$.\\
         \hline
    \end{tabular}
    \caption{Notations used in ILP formulation.}
    \label{tab:ilp-notation}
    \vspace{-3ex}
\end{table}
Table~\ref{tab:ilp-notation} summarizes the notation for the parameters and variables. Given the sets of server instances \( \mathcal{I} \), tasks \( \mathcal{T} \), and instance types \( \mathcal{K} \), the goal is to determine the optimal cluster configuration \( x_{ik} \) and \( y_{i\tau} \) that minimizes the provisioning cost. 

First, it is important to note that both the number of instances to be provisioned and the type of each instance are unknown beforehand. However, an upper bound on the number of instances can be obtained by assigning each task to a separate instance. With this approach, the set of instances \( \mathcal{I} \) is constructed to have cardinality \( |\mathcal{I}| = |\mathcal{T}| \). Additionally, we include a ghost instance type in \(\mathcal{K}\) with zero cost and zero capacity for each type of resource, representing instances that are not provisioned if assigned this type.

Minimizing the total provisioning cost can be formulated as the objective function:
\[
    \min \sum_{i \in \mathcal{I}} \sum_{k \in \mathcal{K}} C_kx_{ik}
\]
This optimization is subject to the following constraints:
\begin{itemize}
    \item Each task is assigned to exactly one instance
    \begin{equation*}
        \sum_{i \in \mathcal{I}} y_{i\tau} = 1,\ \forall \tau \in \mathcal{T}
    \end{equation*}
    \item Each instance is of exactly one instance type
    \begin{equation*}
        \sum_{k \in \mathcal{K}} x_{ik} = 1,\ \forall i \in \mathcal{I}
    \end{equation*}
    \item For each instance, the resource demand does not exceed the resource capacity
    \begin{equation*}
        \sum_{\tau \in \mathcal{T}} D^r_{\tau}y_{i\tau} \leq \sum_{k \in \mathcal{K}} Q^r_kx_{ik},\ \forall i \in \mathcal{I}, r \in \mathcal{R}
    \end{equation*}
\end{itemize}
Solving the above optimization problem for deployment is impractical, as the high computational cost limits its scalability and prohibits it from being employed online. In fact, in the case where there is only a single type of resource (i.e., \(|\mathcal{R}| = 1\)), the problem reduces to VSBPP, which is proven to be NP-hard~\cite{friesen1986vsbpp}. As shown in the micro-benchmark in \S\ref{subsec:scheduling/full-reconfig}, a commercial solver is unable to terminate with an optimal solution within tens of minutes for $200$ tasks and $21$ instance types. Worse, the problem has to be re-solved in each scheduling round whenever a job arrives or completes. 

\subsection{Reservation Price-based Provisioning}
\label{subsec:scheduling/full-reconfig}
\begin{table}[tbp]
    \centering
    \label{tab:rp-example}
    \begin{subtable}[t]{\columnwidth}
        \centering
        \begin{tabular}{|c|c|c|c|c|}
            \hline
            Instance Type & GPU & CPU & RAM (GB) & Cost (\$/hr) \\ \hline
            $it_1$ & 4 & 16 & 244 & 12 \\ \hline
            $it_2$ & 1 & 4 & 61 & 3 \\ \hline
            $it_3$ & 0 & 8 & 32 & 0.8 \\ \hline
            $it_4$ & 0 & 4 & 16 & 0.4 \\ \hline
        \end{tabular}
        \caption{Instance types.}
    \end{subtable}
    \hfill
    \begin{subtable}[t]{\columnwidth}
        \centering
        \begin{tabular}{|c|c|c|c|c|}
            \hline
            Tasks & GPU & CPU & RAM (GB) & \begin{tabular}[c]{@{}c@{}}Reservation \\ Price (\$/hr)\end{tabular} \\ \hline
            $\tau_1$ & 2 & 8 & 24 & 12 \\ \hline
            $\tau_2$ & 1 & 4 & 10 & 3 \\ \hline
            $\tau_3$ & 0 & 6 & 20 & 0.8 \\ \hline
            $\tau_4$ & 0 & 4 & 12 & 0.4 \\ \hline
        \end{tabular}
        \caption{Tasks}
    \end{subtable}
    \caption{Exemplar instance types and tasks.}
    \label{tab: oc-example}
    \vspace{-3ex}
\end{table}
To design an efficient scheduling algorithm, we draw insight from an effective heuristic for VSBPP in one-dimensional space. The heuristic starts by considering the largest bin type and  repeatedly fills the current bin with the largest ball that fits. When no more balls can fit, a new bin of the same type is opened. If the balls in a bin could fit in a smaller bin type, the heuristic switches to the next largest bin type and repeats the process. 

Intuitively, starting with larger bin types increases the likelihood that multiple balls, which might otherwise be assigned to separate smaller bins, can be packed into a single larger bin, thereby reducing the total cost. Similarly, considering balls in descending size minimizes unused space, or fragmentation, within a bin. With divisible bin sizes, the heuristic is proved to have an asymptotic worst-case performance bound of $\frac{11}{9}$~\cite{Kang2003VSBPP}.\\
\textbf{Reservation Price} 
In a multi-dimensional setting, the concept of ``size'' becomes less applicable, as multiple resource types cannot be easily captured by a single dimension. To extend the heuristic for cloud-based cluster scheduling, we need an alternative metric to guide the selection of instance types and tasks while preserving the intuition behind minimizing resource fragmentation. Ideally, the metric should reflect both the quantity and value of the type of resources involved. Since the hourly cost of an instance type is proportional to the amount and type of resources it has, we can evaluate instance types based on their hourly cost. For evaluating tasks, one applicable concept is \textit{reservation price}\footnote{Not to be confused with "reserved instances" in the cloud.}~\cite{steedman2008reservationprice}. In economics, reservation price is the maximum price a buyer is willing to pay for a good or a service. In Eva, the reservation price of executing a task $\tau$, denoted as $RP(\tau)$, is defined as the hourly cost of the cheapest instance type capable of meeting the task's resource demands. In other words, it represents the minimum hourly cost of executing the task on a standalone instance without packing. Consider the example scenario with four instance types and four tasks listed in Table~\ref{tab: oc-example}. The reservation price of tasks \( \tau_1 \), \( \tau_2 \), \(\tau_3\) and \( \tau_4 \) are \$12 (the hourly cost of \( it_1 \)), \$3 (the hourly cost of \( it_2 \)), \$0.8 (the hourly cost of \(it_3\)), and \$0.4 (the hourly cost of \( it_4 \)), respectively.

To determine whether it is cost-efficient to assign a set of tasks to a particular instance, we compare the sum of the reservation prices of the tasks with the hourly cost of the instance. \textit{If the sum of the reservation prices exceeds the instance cost, it indicates that provisioning the instance to host these tasks costs less than assigning each task to its reservation price instance separately.} Using the same example, assigning task \( \tau_1, \tau_2, \) and \( \tau_4 \) to an instance of \( it_1 \) is cost-efficient, as \$12 + \$3 + \$0.4 > \$12. However, assigning only tasks \( \tau_2 \) and \( \tau_4 \) to an instance of \( it_1 \) is not cost-efficient, as \$3 + \$0.4 < \$12. To facilitate discussion, we define the reservation price of a set of tasks \(T\) to be \(RP(T)=\sum_{\tau \in T}RP(\tau)\).

Note that reservation price captures the relative value of different resource types while preserving the ability to co-locate tasks to utilize extra resources. For instance, a CPU task can be assigned to both CPU instances and GPU instances. However, since the reservation price of the CPU task, which is the cost of a CPU instance, is significantly lower than a GPU instance, such an assignment is less likely to be cost-efficient unless there are other GPU tasks being assigned to the same instance.\\
\begin{algorithm}[t]
\caption{Full Reconfiguration}
\label{algo:full-reconfig}
\begin{algorithmic}[1]
\Require $tasks$, $instance\_types$
\State $configuration \gets map()$
\State Sort $instance\_types$ by cost in descending order \label{line:full-reconfig-sort}
\ForAll{$instance\_type$ in $instance\_types$}
    \While{True} \label{line:full-reconfig-it-while-loop-start}
        \State $instance \gets$ new instance of $instance\_type$
        \State $T \gets \{\}$
        \While{tasks can still be packed onto $instance$} \label{line:full-reconfig-job-while-loop-start}
            \State $\tau \gets argmax_{\text{unassigned task } \tau'} RP(T \cup \{\tau'\})$ \label{line:full-reconfig-choose-job}
            \If{$RP(T \cup \{\tau\}) < RP(T)$} \label{line:full-reconfig-job-if-start}
                \State Break
            \EndIf \label{line:full-reconfig-job-if-end}
            \State $T \gets T \cup \{\tau\}$ \label{line:full-reconfig-update-T}
        \EndWhile \label{line:full-reconfig-job-while-loop-end}
        \If{$RP(T) \geq instance.cost$} \label{line:full-reconfig-check-tnoc}
            \State $configuration[instance] \gets T$ \label{line:full-reconfig-worth-it}
        \Else
            \State Break \Comment{Move on to a cheaper instance type} \label{line:full-reconfig-not-worth-it}
        \EndIf
    \EndWhile \label{line:full-reconfig-it-while-loop-end}
\EndFor
\end{algorithmic}
\vspace{-0.5ex}
\end{algorithm}
\textbf{Full Reconfiguration} 
Based on reservation price, we design Eva's scheduling algorithm, shown in Algorithm~\ref{algo:full-reconfig}. We refer to it as the Full Reconfiguration algorithm as it involves considering all the tasks currently in the system for reconfiguration. The algorithm iterates over all available instance types in descending order of cost (Line \ref{line:full-reconfig-sort}). This prioritizes instance types with larger resource capacity and more expensive type of resources, such as GPUs, to minimize costly resource fragmentation. For each instance type, the algorithm repeatedly tries to provision new instances (Line \ref{line:full-reconfig-it-while-loop-start}--\ref{line:full-reconfig-it-while-loop-end}). For a new instance, the algorithm determines the set of tasks \( T \) to assign to this instance through multiple iterations (Line \ref{line:full-reconfig-job-while-loop-start}--\ref{line:full-reconfig-job-while-loop-end}). In each iteration, it selects the unassigned task \( \tau \) that maximizes the total reservation price of \( T \cup \{\tau\} \) (Line \ref{line:full-reconfig-choose-job}). If adding \( \tau \) results in a lower total reservation price, the algorithm stops adding tasks (Line \ref{line:full-reconfig-job-if-start}-\ref{line:full-reconfig-job-if-end}, we explain that this can happen in \S\ref{subsec:scheduling/interference}). Otherwise, \(\tau \) is added to \( T \) (Line \ref{line:full-reconfig-update-T}), and the process continues until no more tasks can be packed onto the instance. The algorithm then checks if assigning \( T \) to the current instance is cost-efficient (Line~\ref{line:full-reconfig-check-tnoc}). If it is, the instance with its assigned tasks \( T \) is added to the new configuration (Line~\ref{line:full-reconfig-worth-it}), and the algorithms tries to provision another instance of the same instance type again. If not, the algorithm moves on to the next, cheaper instance type (Line~\ref{line:full-reconfig-not-worth-it}) and repeats the process.

Note that with Full Reconfiguration, any task-to-instance assignment is guaranteed to be cost-efficient -- the reservation price of the set of tasks assigned to an instance is always at least as high as the instance's actual cost. As a result, while the algorithm prioritizes larger, more costly instance types to reduce resource fragmentation, the cost-efficiency criterion (Line~\ref{line:full-reconfig-check-tnoc}) ensures that such provisioning is justified and avoids unnecessarily leaving resources idle.\\
\textbf{Example} We walk through the execution of the Full Reconfiguration algorithm using the same example provided in Table~\ref{tab: oc-example}. We start by considering the provisioning of an instance from the most expensive instance type \( it_1 \). Task \( \tau_1 \), having the highest reservation price, is assigned to this instance. Similarly, task \( \tau_2 \) is also assigned to the current instance. Task \( \tau_3 \) cannot be accommodated due to insufficient CPU capacity remaining. Moving on, task \( \tau_4 \) is assigned to the current instance. The sum of the reservation price of tasks \( \tau_1 \), \( \tau_2 \), and \( \tau_4 \) is \( \$15.4 \), surpassing the hourly cost of the instance \(\$12\). Thus, this assignment is deemed cost-efficient and is added to the configuration.

Subsequently, another instance of type \( it_1 \) is considered. Task \( \tau_3 \) is attempted to be assigned to this instance. However, since the reservation price of \( \tau_3 \) is less than the hourly cost of the instance, this assignment is not cost-efficient and is discarded. Consequently, we proceed to consider the cheaper instance type \( it_2 \). Similarly, assigning \( \tau_3 \) to an instance of \( it_2 \) is not cost-efficient, so we move on to instance type \( it_3 \).
    
With an instance of type \( it_3 \), the reservation price is equal to the hourly cost of the instance, so we include this assignment in the configuration. As no additional tasks remain, the reconfiguration process concludes. The resulting cluster configuration has an hourly cost of $\$12.8$, which is lower than assigning each task to a separate instance, costing $\$16.2$.\\
\begin{table}[tbp]
    \centering
    \begin{tabular}{l|c|r}
        \hline
        Scheduler & Provisioning Cost & Runtime \\ \hline
        No-Packing & 1.56 \(\pm\) 0.08\(\times\) & 17ms \\
        Full Reconfig. & 1.01 \(\pm\) 0.02\(\times\) & 378ms \\
        ILP & 1\(\times\) & >30min \\ \hline
    \end{tabular}
    \caption{Micro-benchmark results for minimizing provisioning cost. The costs are normalized relative to those incurred by the ILP Scheduler for each trial. Across all 30 trials, the ILP Scheduler timed out with a 30 minutes time limit, and we report the best solution found by then.}
    \label{tab:provisioning-micro-benchmark}
    \vspace{-4ex}
\end{table}
\begin{table}[tbp]
    \centering
    \begin{tabular}{l|cccc}
        \hline
        Num. Tasks & 1000 & 2000 & 4000 & 8000 \\ \hline
        Runtime (sec) & 0.40 & 1.50 & 5.53 & 22.06 \\ \hline
    \end{tabular}
    \caption{Full Reconfiguration runtime.}
    \label{tab:provisioning-scalability}
    \vspace{-5ex}
\end{table}
\textbf{Micro-benchmark} To verify the effectiveness of the Full Reconfiguration algorithm, we conduct a micro-benchmark to measure its ability to minimize the instantaneous provisioning cost given a set of tasks with multi-resource demands. The benchmark consists of 30 independent trials, each involving 200 tasks randomly sampled from the workloads in Table~\ref{tab:evaluate-workload}. The ILP Scheduler is implemented with Gurobi~\cite{gurobi} with a 30 minutes time limit. As shown in Table~\ref{tab:provisioning-micro-benchmark}, the Full Reconfiguration algorithm is able to achieve near-optimal provisioning cost in less than a second.\\
\textbf{Scalability} For each instance, the Full Reconfiguration iterates through all remaining tasks to find a suitable subset for assignment, resulting in a time complexity of $O(|\mathcal{I}||\mathcal{T}|) = O(|\mathcal{T}|^2)$. As shown in Table~\ref{tab:provisioning-scalability}, on a machine with 8 CPU cores, scheduling thousands of tasks takes a few seconds. Scaling beyond this would require reducing the search space of the task set. We plan to study the trade-off between minimizing provisioning cost and scalability in the future.
\\\textbf{Generalizability to Heterogeneous Resources} Different instance families may have varying versions of the same resource type (e.g. A100 vs. V100 GPUs), leading to differences in job throughput across instance families. The concept of reservation price can be extended to account for this, with a slight modification in definition: it can be defined as the minimum cost of executing a single iteration. To evaluate the cost-efficiency of a tasks-to-instance assignment, each task’s reservation price is multiplied by its throughput on the instance family to determine the cost per hour, which is then summed and compared to the hourly instance cost. For simplicity, we use the original definition in the remaining discussion but note that it can be extended to accommodate heterogeneous resources.
\subsection{Incorporating Interference Awareness}
\label{subsec:scheduling/interference}
\textbf{Throughput-Normalized Reservation Price} To account for performance degradation caused by interference among co-located tasks, we extend reservation price to consider throughput. Specifically, if assigning a set of tasks $T$ to an instance results in task $\tau \in T$ having normalized throughput $tput_{\tau, T}$, the \textit{throughput-normalized reservation price} of the task, denoted as $TNRP(\tau, T)$, is defined to be $tput_{\tau, T} \times RP(\tau)$. Intuitively, $TNRP(\tau, T)$ represents the maximum hourly cost the user is willing to pay to host the task $\tau$ at a throughput level of $tput_{\tau, T}$. The throughput could be less than $1$ due to co-location interference. To facilitate discussion, we define the throughput-normalized reservation price of a set of tasks $T$ to be $TNRP(T) = \sum_{\tau \in T} TNRP(\tau, T)$.

A tasks-to-instance assignment is considered cost-efficient if the throughput-normalized reservation price of the set of tasks exceeds the instance's actual cost. Consider the same example in Table~\ref{tab: oc-example}. If co-locating \( \tau_1 \) and \( \tau_2 \) results in normalized throughputs of 0.8 and 0.9, respectively, it is cost-efficient to assign both of them to an instance of \( it_1 \), since \$12 \(\times\) 0.8 + \$3 \(\times\) 0.9 = \$12.3 > \$12. However, if co-locating \( \tau_1 \) and \( \tau_2 \) causes more severe interference, resulting in normalized throughputs of 0.7 and 0.8, respectively, then it is not cost-efficient since \$12 \(\times\) 0.7 + \$3 \(\times\) 0.8 = \$10.8 < \$12.

By replacing \(RP(*)\) with \(TNRP(*)\) in Algorithm~\ref{algo:full-reconfig}, we are able to consider the impact of co-location interference during scheduling. Note that Line~\ref{line:full-reconfig-job-if-start}--\ref{line:full-reconfig-job-if-end} is necessary to ensure that adding more tasks does not result in a decrease in total throughput-normalized reservation price due to severe co-location interference.\\
\textbf{Co-location Throughput Table}
The \texttt{ThroughputMonitor} maintains the co-location throughput table, a data structure that records the throughputs of tasks co-located on the same instance. At every scheduling period, the \texttt{Scheduler} looks up this table to obtain $tput_{\tau, T}$ in order to calculate the throughput-normalized reservation price, which could change as throughput gets updated.

Constructing the table beforehand incurs a high profiling cost that grows exponentially with the number of task types in the system. Instead, Eva builds the co-location throughput table online, updating entries with observed throughput from the tasks. When looking up the co-location throughput of a set of co-located tasks \( T \), the \texttt{ThroughputMonitor} returns the corresponding throughputs if \( T \) has been observed previously and is already recorded in the table. If not, it estimates the throughput of \( \tau \) as \( \prod_{\tau' \in T - \{\tau\}} tput_{\tau, \tau'} \), the product of the pairwise co-location throughputs of task \( \tau \) and the remaining tasks. If \( tput_{\tau, \tau'} \) has not been recorded yet, it is initialized with a default value $t$, which is a tunable parameter of Eva. A smaller $t$ leads to more conservative packing, discouraging the scheduling algorithm from attempting to co-locate tasks. We set $t = 0.95$ in all our experiments.

\subsection{Extending to Multi-Task Jobs}
\label{subsec:scheduling/multi-task}

Up to this point, we have considered each task to be independent. In other words, each task belongs to a single-task job. However, multi-task jobs are prevalent in batch processing. In these cases, the performance of the tasks from the same job $j$ could be interdependent. Specifically, we consider a performance dependency pattern found in data-parallel ML training jobs, where if one task in $j$ experiences performance degradation due to interference from co-location, all tasks in $j$ suffer a decrease in throughput. As a result, treating tasks in a multi-task job as independent tasks in Full Reconfiguration can lead to suboptimal cost-efficiency. To illustrate this, consider a data-parallel ML training job with 4 tasks $\tau_1, \tau_2, \tau_3$ and $\tau_4$. Suppose $\tau_1, \tau_2$ and $\tau_3$ are hosted on individual instances $i_1, i_2$ and $i_3$ without packing, while $\tau_4$ is scheduled to co-locate with other tasks in the system on $i_4$, causing $\tau_4$ to experience co-location interference. While Full Reconfiguration ensures that the throughput-normalized reservation price of the set of tasks assigned to $i_4$ exceeds its cost, the straggler effect of $\tau_4$ leads to reduced throughputs of $\tau_1, \tau_2$ and $\tau_3$, causing the throughput-normalized reservation prices of tasks on $i_1, i_2$ and $i_3$ to be less than the instance cost.

To account for the performance degradation that co-location interference has on a multi-task job, the scheduling algorithm would have to consider the reduction in throughput-normalized reservation price of the entire job, rather than evaluating individual tasks in isolation when assessing cost-efficiency of task-to-instance assignments. Specifically, if assigning a set of tasks $T$ to an instance results in a task $\tau$, which is part of a multi-task job $j$, having normalized throughput $tput_{\tau, T}$, then the throughput-normalized reservation price $TNRP(\tau, T)$ is defined as $RP(\tau) - \sum_{\tau' \in j} (1 - tput_{\tau, T}) \times RP(\tau')$.\\
\textbf{Attributing Source of Interference} For single-task jobs, the co-location throughput table accurately captures co-location interference, as any decrease in a task's throughput can be directly attributed to interference from other tasks sharing the instance. However, for multi-task jobs, a decrease in the throughput of a task \(\tau\) from job \(j\) placed on instance \(i\) can stem from two sources: interference caused by co-located tasks \(T_i\) on instance \(i\), or delays from a straggler task \(\tau'\) of the same job \(j\), which is placed on another instance \(i'\) with co-located tasks \(T_{i'}\). In the latter case, naïvely recording \(tput_{\tau, T_i}\) in the co-location throughput table may lead to overly pessimistic attribution of co-location interference, resulting in conservative packing decisions in Full Reconfiguration.

To address this issue, the \texttt{ThroughputMonitor} uses a set of rules to logically deduce the source of interference. Suppose we have a multi-task job \(j\), consisting of tasks \(\tau_1, \tau_2, ..., \tau_n\), each placed on instances \(i_1, i_2, ..., i_n\) alongside co-located tasks \(T_1, T_2, ..., T_n\), respectively. When the throughput of job $j$ is observed, the \texttt{ThroughputMonitor} attempts to identify the straggler and updates only a single entry $tput_{\tau, T}$, following these rules:
\vspace{-1ex}
\begin{itemize}
    \item No previous observations: If none of \(tput_{\tau_1, T_1}\), \(tput_{\tau_2, T_2}\), ..., \(tput_{\tau_n, T_n}\) has been recorded, the table updates the entry for the task \(\tau\) co-located with the most tasks \(T\).
    \item Some previous observations with lower throughput: If any previously recorded throughput is lower than the currently observed value, the table updates the entry for the task \(\tau\) co-located with tasks \(T\) that had the lowest recorded throughput.
    \item All previous observations have higher throughput: If all previous recorded throughput show higher throughput than the current observation, the table updates the entry for the unrecorded task \(\tau\) co-located with the most tasks \(T\).
\end{itemize}
By following these rules, the recorded throughput in the co-location throughput table is guaranteed to represent a lower bound of the actual co-location throughput. As more observations are made, the table is updated, and the recorded values are adjusted upwards, reflecting a more accurate estimation of the true co-location interference.

Eva’s approach to managing multi-task jobs assumes a performance dependency pattern found in data-parallel jobs, where all tasks are interdependent. Extending this to accommodate more general dependency patterns is left for future work.\\
\begin{table}[tbp]
    \centering
    \begin{tabular}{l|c|c}
        \hline
        Scheduler & Norm. Total Cost & JCT (hours) \\ \hline
        No-Packing & 100\% & 4.44 \(\pm\) 0.35\\
        \texttt{Eva-Single} & 79.5\% \(\pm\) 3.8\% & 5.11 \(\pm\) 0.51 \\
        \texttt{Eva-Multi} & 74.2\% \(\pm\) 4.2\% & 4.55 \(\pm\) 0.37\\ \hline
    \end{tabular}
    \caption{Micro-benchmark results for scheduling multi-task jobs. The costs are normalized relative to those incurred by the No-Packing Scheduler for each trial.}
    \vspace{-4ex}
    \label{tab:multi-task-micro-benchmark}
\end{table}
\textbf{Micro-benchmark} To verify the effectiveness of our extension for multi-task jobs, we conduct simulations using our simulator (\S\ref{sec:implementation}). The simulation includes 10 independent trials, with each trial involving the scheduling of 100 multi-task jobs arriving over time. Each job consists of 4 identical tasks, uniformly sampled from Table~\ref{tab:evaluate-workload}, and has a job duration ranging from 0.5 to 16 hours. Table~\ref{tab:multi-task-micro-benchmark} shows the result of Eva with (\texttt{Eva-Multi}) and without (\texttt{Eva-Single}) considering the interdependency of tasks within a multi-task job. While both schedulers substantially reduce the total cost due to the effectiveness of Full Reconfiguration, jobs in \texttt{Eva-Multi} have lower JCT, reflecting their reduced impact from degrading throughput caused by a single interfered task, which further lowers the total cost.

\subsection{Migration Awareness}
\label{subsec:scheduling/migration}
\textbf{Partial Reconfiguration} The Full Reconfiguration algorithm holistically optimizes provisioning cost by considering all the tasks in the system for reconfiguration. However, it does not take the current cluster configuration into account, which might lead to excessive task migrations and frequent instance launches or terminations when switching from one configuration to another. To mitigate this, we introduce a reconfiguration scheme that only considers a subset of tasks for reconfiguration, leaving the rest of the cluster configuration unchanged. We refer to this heuristic as the Partial Reconfiguration algorithm. Specifically, the subset of tasks consists of tasks from recently submitted jobs that have not yet been assigned to any instances, and existing tasks on instances that are no longer considered cost-efficient. The latter occurs when the throughput-normalized reservation price of the tasks on an instance drops below the instance's hourly cost. This decrease can result from job completion or reduced throughput due to co-location interference. The subset of tasks is processed using Algorithm~\ref{algo:full-reconfig} to obtain an updated configuration. Combined with the unchanged configuration of the remaining tasks and instances, this becomes the output configuration of Partial Reconfiguration.\\
\textbf{Full Reconfiguration vs. Partial Reconfiguration} The two reconfiguration algorithms prioritize maximizing provisioning cost-efficiency and minimizing migration overhead, respectively. 
Using either one alone is insufficient: Full Reconfiguration at every scheduling period incurs significant migration overhead, while Partial Reconfiguration deviates from the optimal configuration that minimizes provisioning cost over time.
As a result, Eva takes an ensemble approach. At each scheduling period, Eva runs both algorithms to obtain two configurations and decides which one to adopt. Intuitively, Full Reconfiguration is preferred if its configuration yields significant provisioning cost savings that justify the incurred migration overhead. However, provisioning cost savings depend not only on the instances provisioned but also on how long the configuration will last. If a job is submitted or completes shortly after a Full Reconfiguration, triggering another Full Reconfiguration that again involves migrating a lot of tasks, the initial reconfiguration provides little cost benefit and may even result in extra costs due to the incurred migration overhead. In such cases, it is better to adopt Partial Reconfiguration, accepting a suboptimal cluster configuration in terms of provisioning cost and wait for the job arrival or completion that triggers the Full Reconfiguration. The main challenge here is that we do not know when the next job arrival or completion will happen, and whether they will trigger a Full Reconfiguration.

We propose a quantitative criterion for deciding between Full Reconfiguration and Partial Reconfiguration. Let \( S_F \) (\( S_P \)) be the instantaneous provision cost saving of the Full (Partial) Reconfiguration, which is calculated as the sum of the differences between the the throughput-normalized reservation price and the actual cost of each instance. Let \( M_F\) (\(M_P\)) be the migration cost incurred by  Full (Partial) Reconfiguration, which is calculated based on task migration delays and the cost of the involved instances. Let \( D\) be the duration of the new configuration, i.e., the length of time the new configuration will last until the next Full Reconfiguration. If we were able to know \( D\), the criterion would be to choose Full Reconfiguration if 
\begin{equation}
    S_F \times D - M_F > S_P \times D - M_P
    \label{eqn:full_vs_partial}
\end{equation}
However, \( D \) is unknown in advance as we have discussed. To estimate \( D \), we first note that Full Reconfiguration can only happen when a job arrives or completes. Otherwise, the configuration will not change as the set of tasks remains the same. We refer to job arrivals or completions as ``events.'' Assume that the occurrence of these events follows a Poisson process with a rate of $\lambda$. Let $N(x)$ be the number of events that happens between time $[0, x]$. Let $p$ be the probability that an event triggers a Full Reconfiguration. Assuming independence, the probability distribution of the number of events until the next Full Reconfiguration can be modeled as a geometric distribution with parameter $p$. Therefore, the probability that the next Full Reconfiguration will happen between time $[0, x]$ can be estimated as $F(x) = 1 - (1 - p)^{E[N(x)]} = 1 - (1-p)^{\lambda x} $. Similar to calculating the mean time to failure~\cite{rausand2003mttr}, we can calculate the mean time to the next Full Reconfiguration as 
\begin{equation*}
    \hat{D} = \int_0^{\infty} (1-F(x))dx = \int_0^{\infty} (1-p)^{\lambda x}dx = -\frac{1}{\lambda \ln(1-p)},
\end{equation*}
which could be used as an estimate of \( D \) for calculating Equation~\ref{eqn:full_vs_partial}. Note that $\lambda$ and $p$ can be empirically estimated in the system.
\section{Implementation}
\label{sec:implementation}
We implemented Eva and a simulator in Python with approximately 5,700 lines of code. Eva follows a modular architecture (\S\ref{sec:design}) for extensibility and adopts a centralized master-worker model. The master manages cloud instances through existing cloud platform APIs. Once an instance is instantiated, a worker is launched on the instance, which communicates with the master through gRPC~\cite{grpc}. To use Eva, users simply provide a Dockerfile with their execution artifacts and specify the required resources, similar to existing container-based cloud platforms such as Amazon Elastic Kubernetes Service~\cite{amazoneks}, Azure Kubernetes Service~\cite{azureks}, and Google Kubernetes Engine~\cite{gke}.\\
\textbf{Task Execution and Submission} Tasks are executed as Docker containers to ensure portability and environment isolation. Users submit jobs to Eva by specifying the Dockerfile and the resource demand vector \([g, c, m]\) for each task, which details the required amounts of GPU, CPU, and RAM for task execution. To leverage heterogeneous cloud instances, users can specify multiple resource demand vectors for different instance types. For example, a task could have demand vectors [0, 8, 8] for P3 instance types and [0, 4, 8] for C7i instance types. All instances in the cluster have access to a global storage. The workers mount the global storage to the Docker containers, so that each task can access necessary artifacts such as datasets and checkpoints from it.\\
\textbf{Throughput Monitoring} To facilitate throughput monitoring, the worker communicates with each job via \texttt{EvaIterator}, a lightweight API wrapper around common data iterators. At the start of each scheduling round, the worker requests the throughput over a user-specified time window (e.g., the last 10 minutes) from each job and reports this information back to the master. This data is used to update the co-location throughput table.\\
\textbf{Simulator} We implemented a simulator to facilitate the design and evaluation of Eva. During simulation, Eva operates as it would in a real-world deployment, but interacts with a simulated cloud environment.
The simulator reads a workload trace and notifies Eva of job arrivals and their resource demands. Given the jobs in the cluster, the \texttt{Scheduler} determines the cluster configuration. Based on this configuration, the \texttt{Provisioner} and the \texttt{Executor} issue operational commands -- such as launching or terminating cloud instances and migrating tasks between instances -- to the simulated cloud environment. To model real-world cloud behavior, the simulator incorporates operation delays measured from cloud instances (Table~\ref{tab:delay}), which affect instance uptime and thus overall provisioning cost. Job execution and progress are also simulated using real-world throughput data (Figure~\ref{fig:tput-matrix}). The throughput of a task is changed over time based on task co-location to account for co-location interference, with the interference data drawn from our measurements. Eva's scheduler is not provided with this data but observes task throughput and interference through the \texttt{ThroughputMonitor}, which interacts with the simulator to periodically collect throughput from tasks.
\vspace{-2ex}
\section{Evaluation}
\label{sec:eval}
We evaluate Eva on AWS EC2 with synthetic traces that consist of batch processing jobs from a wide range of applications. To test Eva's effectiveness on a larger scale, we run simulated experiments with production cluster traces.
\subsection{Experiment Setup}
\label{subsec:eval/setup}
\begin{table*}[tbp]
    \centering
    \settowidth\tymin{\textbf{A}}
    \begin{tabulary}{\textwidth}{LlLCCCCC}
        \hline
         \multirow{2}{*}{Workload} & \multirow{2}{*}{Description} & \multirow{2}{*}{Dataset} & \multicolumn{3}{c}{Resource Demand} & \multicolumn{2}{c}{Mig. Delay (sec)}\\
         \cline{4-6}\cline{7-8}
         & && GPU & CPU & RAM (GB) & Checkpoint & \mbox{Launch}\\
         \hline
         ML -- Image Classification & ResNet18~\cite{he2016resnet}--2 Tasks & ImageNet~\cite{russakovsky2015imagenet} & 1 & 4 & 24 & 2 & 80\\
         ML -- Image Classification & ResNet18~\cite{he2016resnet}--4 Tasks & ImageNet~\cite{russakovsky2015imagenet} & 1 & 4 & 24 & 2 & 80\\
         ML -- Image Classification & ViT~\cite{kolesnikov2021vit} & ImageNet~\cite{russakovsky2015imagenet} & 2 & 8 & 60 & 3 & 143 \\
         ML -- I2I Translation & CycleGAN~\cite{zhu2017cyclegan} & monet2photo~\cite{zhu2017cyclegan} & 1 & 4 & 10 & 7 & 2\\
         ML -- Language Modeling & GPT2~\cite{radford2019gpt2} & WikiText-2~\cite{merity2017wikitext2} & 4 & 4 & 10 & 30 & 15 \\
         ML -- Graph Embedding & \mbox{GraphSAGE}~\cite{hamilton2017graphsage} & ogbn-products~\cite{hu2020ogb} & 1 & 8 & 50 & 2 & 160\\
         ML -- Graph Embedding & GCN~\cite{kipf2017gcn} & ogbn-products~\cite{hu2020ogb} & 0 & \mbox{12 (6)} & 40 & 2 & 28 \\
         ML -- RL & A3C~\cite{mnih2016a3c} & Pong & 0 & \mbox{10 (4)} & 8 & 2 & 10\\
         BioInfo -- Sequence Alignment & Diamond~\cite{buchfink2021diamond} & UniRef50 \& UniProtKB/Swiss-Prot~\cite{uniprot} & 0 & \mbox{14 (8)} & 16 & 8 & 12\\
         Physics -- Computational Fluid Dynamics & \mbox{OpenFOAM}~\cite{weller1998openfoam} & Motorbike & 0 & \mbox{8 (6)} & 8 & 21 & 1\\
         \hline
    \end{tabulary}
    \caption{Evaluated workloads and resource demand per task. All workloads are single-task jobs except for ResNet18. For CPU demand, the number outside the parentheses represents the demand on P3 instances, while the number in parentheses (when present) represents the demand on C7i and R7i instances. Since C7i and R7i instances have CPUs with higher frequency, CPU jobs can achieve the same throughput on these instances with fewer CPUs.}
    \label{tab:evaluate-workload}
    \vspace{-3ex}
\end{table*}
\textbf{Cloud Infrastructure} Our experiments consider 21 instance types from 3 families on AWS EC2: P3 instances (GPU instances), C7i instances (compute-optimized instances), and R7i instances (memory-optimized instances). All instances are provisioned in the same region. If an instance type is not available in the default availability zone, the \texttt{Provisioner} retries in other availability zones until an instance is successfully provisioned. An S3 bucket is used as the global storage.\\
\begin{table}[tbp]
    \centering
    \small
    \begin{tabular}{l|ccccc}
        \hline
        GPU Demand & 0 & 1 & 2 & 4 & 8\\ \hline
        Job Population & 13.41\% & 86.17\% & 0.20\% & 0.18\% & 0.04\%\\\hline
    \end{tabular}
    \caption{Alibaba trace job composition by GPU demands.}
    \label{tab:alibaba-trace-job-composition}
    \vspace{-4ex}
\end{table}
\begin{table}[tbp]
    \centering
    \small
    \begin{tabular}{l|cccc}
        \hline
        & Mean (hr) & Median (hr) & P80 (hr) & P95 (hr)\\\hline
        Alibaba~\cite{weng2023trace} & 9.1 & 0.2 & 1.0 & 5.2 \\ 
        Gavel~\cite{narayanan2020gavel} & 16.7 & 4.5 & 16.4 & 96.6 \\\hline
    \end{tabular}
    \caption{Job duration in simulation experiments.}
    \label{tab:simulation-trace-job-duration}
    \vspace{-4ex}
\end{table}
\textbf{Workloads and Traces} Our experiments considers 10 different batch processing workloads from a variety of ML and scientific computation applications, as shown in Table~\ref{tab:evaluate-workload}.

For the physical experiments, we generate synthetic traces similar to prior work~\cite{zheng2023shockwave}. We conduct two physical experiments at different scales. The small-scale experiment uses a trace with 32 jobs, while the large-scale experiment uses a trace with 120 jobs. These jobs are sampled from the 10 workloads in Table~\ref{tab:evaluate-workload}. The job durations range from 0.5 to 3 hours long, and the job arrival times are generated according to a Poisson arrival process with an average inter-arrival time of 20 minutes.

For the simulated experiments, we use the publicly available production trace (cluster-trace-gpu-v2023) from Alibaba \cite{weng2023trace}, which captures the usage patterns of Alibaba's internal batch-job users. We preserve the resource demands for GPU, CPU, and RAM for each task. The original trace consists only of single-task jobs. To maintain the integrity of the trace, we treat each task as a single-task job in our simulation experiments. However, in \S\ref{subsec:eval/multi-task}, we present an experiment extending the trace to include multi-task jobs. The job composition of the trace in terms of GPU demand is shown in Table~\ref{tab:alibaba-trace-job-composition}. After removing failed jobs and jobs that have resource demands that cannot be accommodated by any of the 21 instance types, the final trace consists of 6,274 jobs. For job duration, we consider the two cases shown in Table~\ref{tab:simulation-trace-job-duration}. The original trace includes a high proportion of short jobs, with 80\% lasting less than an hour and half lasting less than 11 minutes. To better represent the long-running nature of ML training jobs, we also use the job duration modeling approach from Gavel~\cite{narayanan2020gavel} in separate experiments: each job duration is sampled from an exponential distribution, with the duration set to \( 10^x \) minutes, where \(x\) is drawn uniformly from \([1.5, 3]\) with 80\% probability, and from \([3, 4]\) with 20\% probability. Job arrival times are generated following the same procedure as in the physical experiment and we also study the effect of varying the job arrival rate in \S\ref{subsec:eval/arrival}. We assign each job a workload from Table~\ref{tab:evaluate-workload} to simulate the job's migration overhead and co-location throughput based on those of the associated workloads.\\ 
\textbf{Baselines} We compare Eva against the following schedulers, which represent either the state-of-the-art or the most commonly used solutions for hosting jobs in cloud environments: 
\vspace{-5ex}
\begin{itemize}
    \item No-Packing Scheduler: Each task is hosted on a separate instance without any co-location. As a result, tasks do not experience interference from other tasks. This is representative of the strategy adopted by the majority of existing cloud-based cluster managers~\cite{yang2023skypilot, harlap2017proteus, shastri2017hotspot, xu2019ispot}.
    \item Stratus~\cite{chung2018stratus}: Stratus minimizes task migration overhead by co-locating tasks with similar finish times. To achieve this, it relies on job runtime estimates. For comparison against Stratus's best-case scenario, we provide Stratus with the job duration calculated as the total iterations divided by the throughput.
    \item Synergy~\cite{mohan2022synergy}: Synergy employs a best-fit packing heuristic to minimize resource fragmentation in a fixed-sized cluster. We adapt Synergy for cloud-based clusters with variable size by launching the lowest-cost instance type capable of accommodating a task when no existing instance in the cluster has enough capacity. In addition, we enhance the heuristic to be interference-aware by incorporating throughput-normalized reservation price when assigning tasks to existing instances.
    \item Owl~\cite{tian2022owl}: Owl minimizes interference by only co-locating task pairs that result in low interference, with its scheduling algorithm prioritizing co-locations that maximizes resource allocation. It relies on profiling the co-location throughput for all task pairs in advance, and we provide this profile exclusively to Owl. Additionally, we extend Owl's scheduling algorithm to optimize for cost-efficiency by considering task pairs in descending ratio of their throughput-normalized reservation price to the cost of the least expensive instance type that can accommodate them.
\end{itemize}
\vspace{-1ex}
\noindent\textbf{Metrics} We report the total cost incurred by each scheduler. For meaningful comparison across traces, we show the normalized cost of each scheduler, calculated relative to No-Packing Scheduler's cost of each trace. Additionally, we include metrics such as resource allocation (the ratio of allocated resources to total resources), normalized job throughput and JCT to provide a comprehensive understanding of the factors contributing to cost reduction.
\subsection{End-to-End Physical Experiment Results}
\label{subsec:eval/physical}
\begin{table*}[t]
    \centering
    \begin{minipage}{0.6\textwidth}
        \centering
        \begin{tabular}{llrrcccc}
            \hline
            \multirow{2}{*}{Scheduler} & \multicolumn{1}{c}{Total Cost} & Instances & Migration & \multicolumn{3}{c}{Avg. Resource Alloc.} \\
            \cline{5-7}
             & \multicolumn{1}{c}{(Norm. Cost)} & Launched & per Task & GPU & CPU & RAM \\
            \hline
            No-Packing & \$536.07 (100\%)  & 126 & 0 & 67\% & 77\% & 28\% \\
            Stratus & \$533.62 (99.5\%) & 76 & 0.02 & 64\% & 72\% & 31\% \\
            Eva & \$452.40 (84.4\%) & 154 & 1.23 & 76\% & 85\% & 41\% \\
            \hline
        \end{tabular}
        \caption{End-to-end physical experiment with 120 jobs.}
        \label{tab:physical-results-large}
    \end{minipage}%
    \hfill
    \begin{minipage}{0.35\textwidth}
        \centering
        \includegraphics[width=0.65\textwidth]{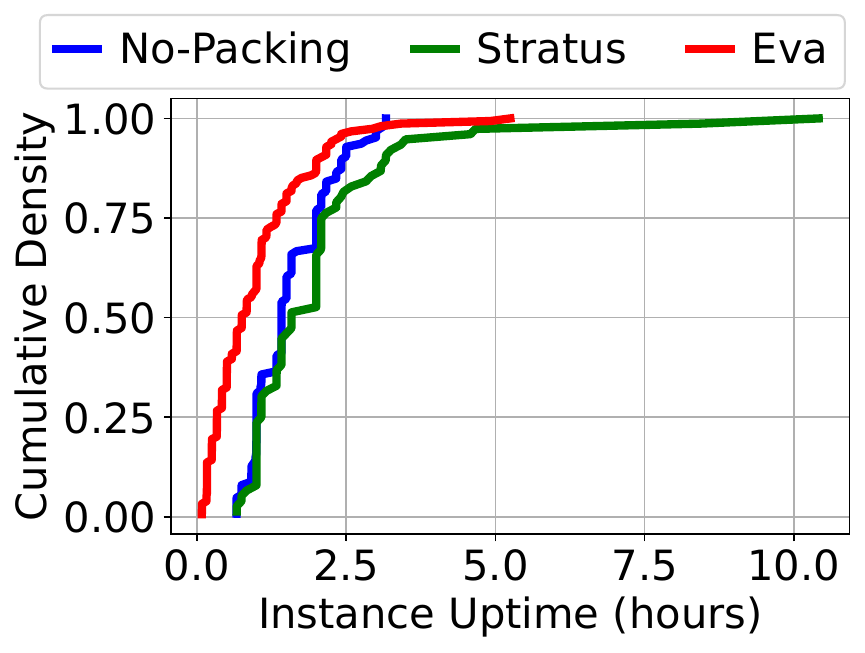}
        \vspace{-2ex}
        \captionof{figure}{Instance uptimes with 120 jobs.}
        \label{fig:physical_instance_utime_cdf}
    \end{minipage}
\end{table*}

\begin{table}[t]
    \centering
    \begin{tabular}{llcccc}
        \hline
        \multirow{2}{*}{Scheduler} & \multicolumn{1}{c}{Total Cost} & \multicolumn{3}{c}{Avg. Resource Alloc.} \\
        \cline{3-5}
         & \multicolumn{1}{c}{(Norm. Cost)} & GPU & CPU & RAM \\
        \hline
        No-Packing & \$163.87 (100\%) & 63\% & 76\% & 29\% \\
        Stratus & \$145.76 (88.9\%)& 67\% & 74\% & 32\% \\
        Synergy & \$145.80 (89.0\%) & 66\% & 86\% & 32\% \\
        Owl & \$143.75 (87.7\%) & 69\% & 88\% & 38\% \\
        Eva & \$123.03 (75.1\%) & 76\% & 89\% & 42\% \\
        \hline
    \end{tabular}
    \caption{End-to-end physical experiment with 32 jobs.}
    \label{tab:physical-results}
    \vspace{-1ex}
\end{table}
\begin{table}[h]
    \centering
    \begin{tabular}{lccS[table-format=0.1\%]}
        \hline
        Scheduler & Actual Cost & Simulated Cost & {Difference}\\
        \hline
        No-Packing & \$163.87 & \$160.74 & -1.9\%\\
        Stratus & \$145.76 & \$152.94 & 4.9\%\\
        Synergy & \$145.80 & \$141.17 & -3.2\%\\
        Owl & \$143.75 & \$146.84 & 2.2\%\\
        Eva & \$123.03 & \$123.78 & 0.6\%\\
        \hline
    \end{tabular}
    \caption{Simulator fidelity.}
    \label{tab:simulator-fidelity}
    \vspace{-3ex}
\end{table}
Table~\ref{tab:physical-results-large} and Figure~\ref{fig:physical_instance_utime_cdf} show the results of the physical experiment conducted with the 120-job trace, using the No-Packing scheduler (the most common solution), Stratus (state-of-the-art for cloud-based cluster scheduling), and Eva. Eva reduces the total cost by 15\% compared to the baselines. In contrast to the baseline schedulers, Eva actively adjusts the cluster configuration through selecting suitable instance types and migrating tasks, resulting in more instances launched over time, more migration per task (Table~\ref{tab:physical-results-large}), and shorter uptime per instance (Figure~\ref{fig:physical_instance_utime_cdf}). This minimizes resource fragmentation and addresses the mismatch between resource demand and instance capacity, resulting in the highest cluster-wide resource allocation across all three types of resources.

Table~\ref{tab:physical-results} shows the results of the physical experiment conducted with the 32-job trace using all baseline schedulers. Similar to the larger trace, Eva reduces the total cost by 15-25\% compared to existing baselines. In addition, we run the same trace using our simulator and compare the simulated results to the observed results from the physical experiment. As shown in Table~\ref{tab:simulator-fidelity}, the difference between the total cost in simulated and physical experiments is within 5\%, indicating the high fidelity of the simulator.
\subsection{End-to-End Simulation Results}
\label{subsec/simulation}
\newcolumntype{b}{>{\hsize=1.2\hsize}Y}
\newcolumntype{m}{>{\hsize=.8\hsize}X}
\newcolumntype{a}{>{\hsize=.6\hsize}Y}
\newcolumntype{s}{>{\hsize=.5\hsize}Y}
\begin{table*}[tbp]
    \centering
    \begin{tabular}{llcccc}
        \hline
        \multirow{2}{*}{Scheduler} & \multicolumn{1}{c}{Total Cost} & Num. of Tasks & Norm. & JCT & Job Idle Time \\
         & \multicolumn{1}{c}{(Norm. Cost)} & per Instance & Job Tput & (hours) & (hours) \\
        \hline
        \text{No-Packing} & \$480,130 (100\%) & 0.99 & 1 & 9.18 & 0.10\\
        Stratus & \$344,171 (72\%) & 1.60 & 0.94 & 9.71 & 0.05\\
        Synergy & \$368,033 (77\%) & 1.72 & 0.93 & 9.68 & 0.05\\
        Owl & \$376,678 (78\%) & 1.81 & 0.96 & 9.84 & 0.16\\
        Eva & \$289,908 (60\%) & 2.05 & 0.91 & 10.55 & 0.11\\
        \hline
    \end{tabular}
    \caption{End-to-end simulation with Alibaba job duration. Job idle time represents the duration a job is not executing due to delays shown in Table~\ref{tab:delay}.}
    \label{tab:simulation-results-pai-duration}
    \vspace{-5ex}
\end{table*}
\begin{table*}[tbp]
    \centering
    \begin{tabular}{llcccc}
        \hline
        \multirow{2}{*}{Scheduler} & \multicolumn{1}{c}{Total Cost} & Num. of Tasks & Norm. & JCT & Job Idle Time \\
         & \multicolumn{1}{c}{(Norm. Cost)} & per Instance & Job Tput & (hours) & (hours) \\
        \hline
        \text{No-Packing} & \$831,227 (100\%) & 1 & 1 & 16.81 & 0.10\\
        Stratus & \$560,067 (67\%) & 2.28 & 0.90 & 18.89 & 0.05\\
        Synergy & \$556,901 (67\%) & 2.26 & 0.89 & 19.03 & 0.06\\
        Owl & \$629,673 (75\%) & 1.84 & 0.94 & 18.05 & 0.19\\
        Eva & \$483,472 (58\%) & 2.59 & 0.89 & 19.42 & 0.17\\
        \hline
    \end{tabular}
    \caption{End-to-end simulation with Gavel job duration.}
    \label{tab:simulation-results}
    \vspace{-3ex}
\end{table*}
To validate Eva's benefits in larger scale and more realistic setting, we run simulated experiments using the Alibaba production trace that consists of 6,274 jobs. The results are shown in Table~\ref{tab:simulation-results-pai-duration} and Table~\ref{tab:simulation-results}. In line with the physical experiments, Eva has the lowest total cost among the five schedulers, reducing the cost by 13-42\%. We observe similar pattern of resource allocation as the physical experiments. Compared to other packing schedulers, Eva's interference-aware scheduling packs more tasks per instance and achieves higher resource allocation while maintaining similar job throughput. Although more aggressive reconfiguration incurs higher idle time, Eva makes this trade-off to achieve better resource allocation and lower total cost. In addition, compared to the No-Packing Scheduler, Eva and other packing schedulers experience a 5–16\% increase in JCT, primarily due to decreased throughput from co-location. Since our main objective is to minimize the overall costs, this trade-off is considered worthwhile for a 42\% cost reduction. We plan to study how the scheduling algorithm can be extended to consider JCT as part of the objective in the future.

\subsection{Impact of Co-location Interference}
\label{subsec:eval/interference}
\begin{figure}[t]
    \centering
    \includegraphics[width=\linewidth]{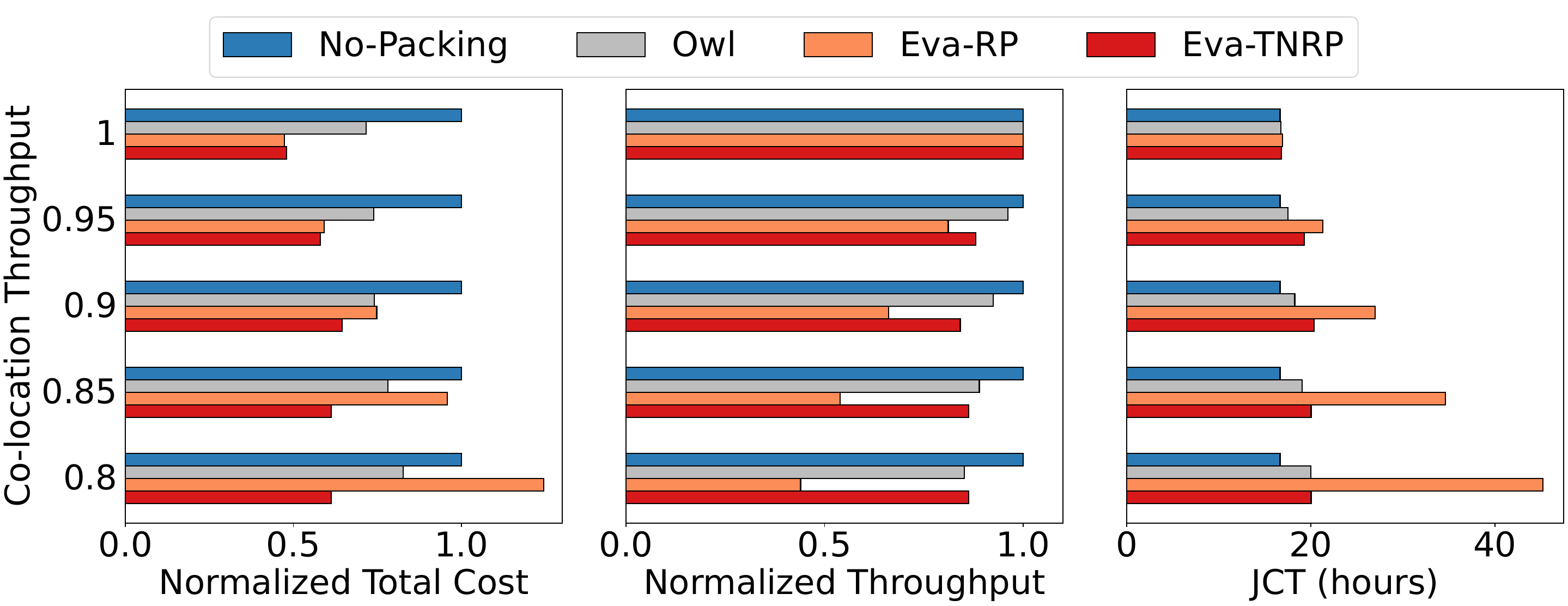}
    \caption{Impact of co-location interference.}
    \label{fig:vary_interference}
\end{figure}
To reinforce the importance of considering co-location interference in scheduling, we run simulated experiments with different degree of interference when jobs are co-located on the same instance. Specifically, we run the Alibaba trace with simulated pairwise co-location throughput set to \{1, 0.95, 0.9, 0.85, 0.8\}. For example, if pairwise co-location throughput is set to 0.9, then when two jobs are co-located, they both have normalized throughput of 0.9. We compare Eva with and without considering interference: \texttt{Eva-TNRP} and \texttt{Eva-RP}. \texttt{Eva-TNRP} uses throughput-normalized reservation price in Algorithm~\ref{algo:full-reconfig}, while \texttt{Eva-RP} uses reservation price. We also include two baseline schedulers that prioritize minimizing co-location interference -- No-Packing Scheduler and Owl.

Figure ~\ref{fig:vary_interference} illustrates that as the degree of interference increases (i.e., as co-location throughput decreases), \texttt{Eva-RP} experiences a significant decrease in job throughput, leading to an increase in JCT. Consequently, while packing improves resource allocation, the longer job runtime necessitates longer instance provisioning, resulting in increased total cost. Conversely, accounting for throughput degradation when evaluating cost-efficiency in scheduling, \texttt{Eva-TNRP} maintains a throughput level similar to Owl, which is designed to minimize co-location interference. This, combined with higher resource allocation from task packing, enables \texttt{Eva-TNRP} to reduce the overall cost even in scenarios with high degrees of interference. We note that in extreme cases where severe interference makes any packing sub-optimal, Eva refrains from co-locating tasks, reducing to No-Packing Scheduler.
\subsection{Impact of Migration Overhead}
\label{subsec:eval/migration}
\begin{figure}[t]
    \centering
    \begin{subfigure}[t]{\columnwidth}
        \centering
        \includegraphics[width=0.93\columnwidth]{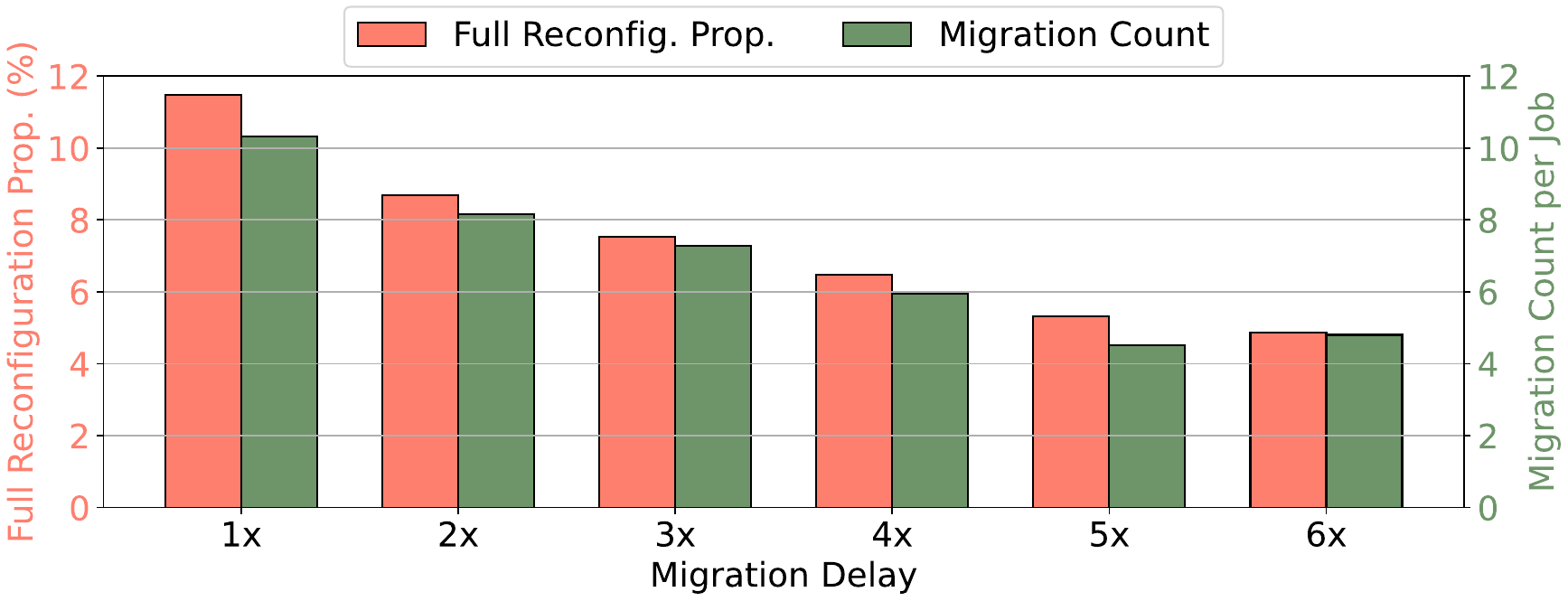}
        \caption{Full reconfiguration proportion.\label{fig:vary_migration_overhead_global_reconfig_prop}}
    \end{subfigure}
    \begin{subfigure}[t]{\columnwidth}
        \centering
        \includegraphics[width=0.93\columnwidth]{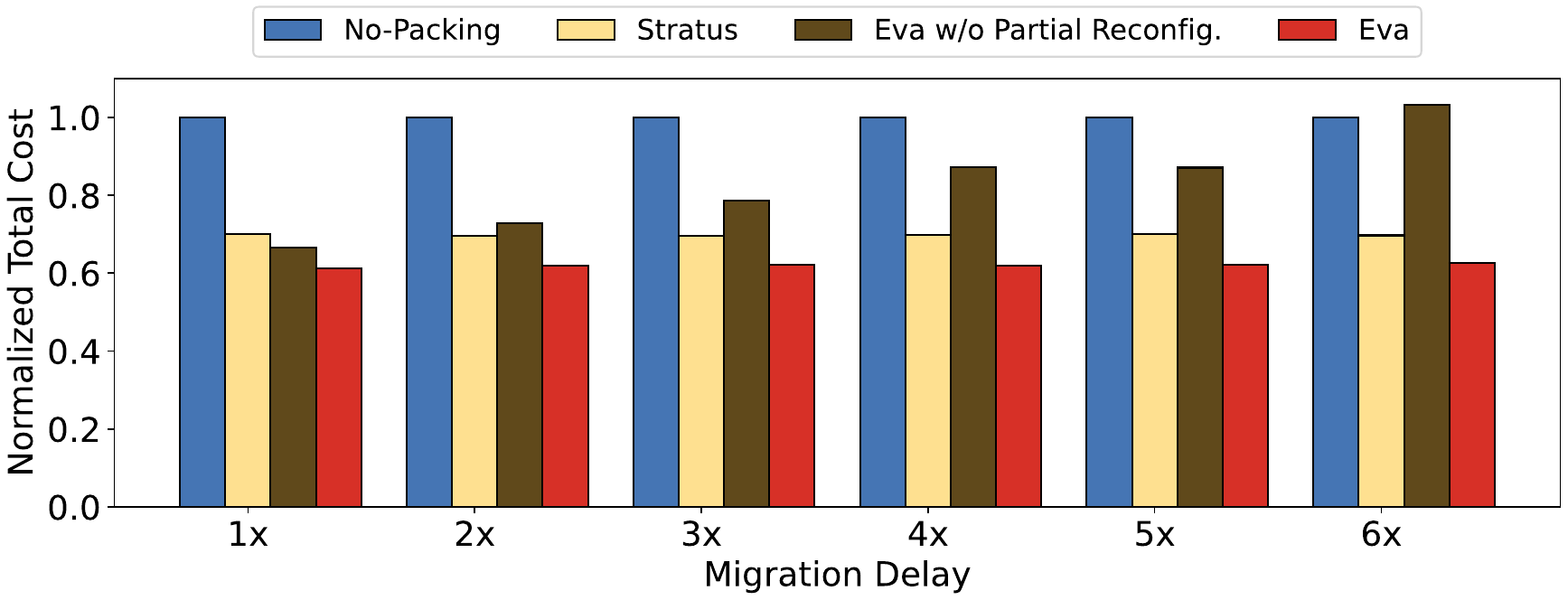}
        \caption{Total cost.\label{fig:vary_migration_overhead_total_cost}}
    \end{subfigure}
    \vspace{-2ex}
    \caption{Impact of migration overhead. \(2 \times\) means each job's migration delay is set to twice its original delay duration.}
    \label{fig:vary_migration_overhead}
    \vspace{-1ex}
\end{figure}
As ML models grow in size, it becomes more expensive to migrate them between instances. To better understand how Eva's ensembling approach handles trade-off between migration overhead and provision savings in these scenarios, we run the same Alibaba trace with varying levels of simulated job migration delay. Figure~\ref{fig:vary_migration_overhead_global_reconfig_prop} shows the proportion of Full Reconfiguration adopted as the final configuration (left y-axis) and the migration count per job (right y-axis) of Eva under various levels of migration delay. Since Full Reconfiguration prioritizes minimizing provisioning costs at the expense of increased job migrations, significant migration overhead can overshadow the provisioning savings when migration delay increases. In such cases, Full Reconfiguration becomes less likely to be adopted because the increased \( M_F \) in Equation~\ref{eqn:full_vs_partial} makes it less likely to hold. Instead, Partial Reconfiguration, which maintains the majority of current cluster configuration and only migrates a small subset of essential jobs, is more likely to be adopted, resulting in a decrease in migration count per job.

Figure~\ref{fig:vary_migration_overhead_total_cost} shows that using Full Reconfiguration alone without Partial Reconfiguration results in a noticeable increase in total cost, which becomes more pronounced as migration delays increase. On the other hand, baseline schedulers like Stratus, which prioritize minimizing migration, remains largely unaffected. By balancing provisioning savings and migration overhead, Eva's ensembling approach allows for significant cost reductions to be maintained even in the presence of substantial job migration delays. 

\subsection{Impact of Workload Composition}
\label{subsec:eval/composition}
\begin{figure}[t]
    \centering
    \includegraphics[width=0.92\linewidth]{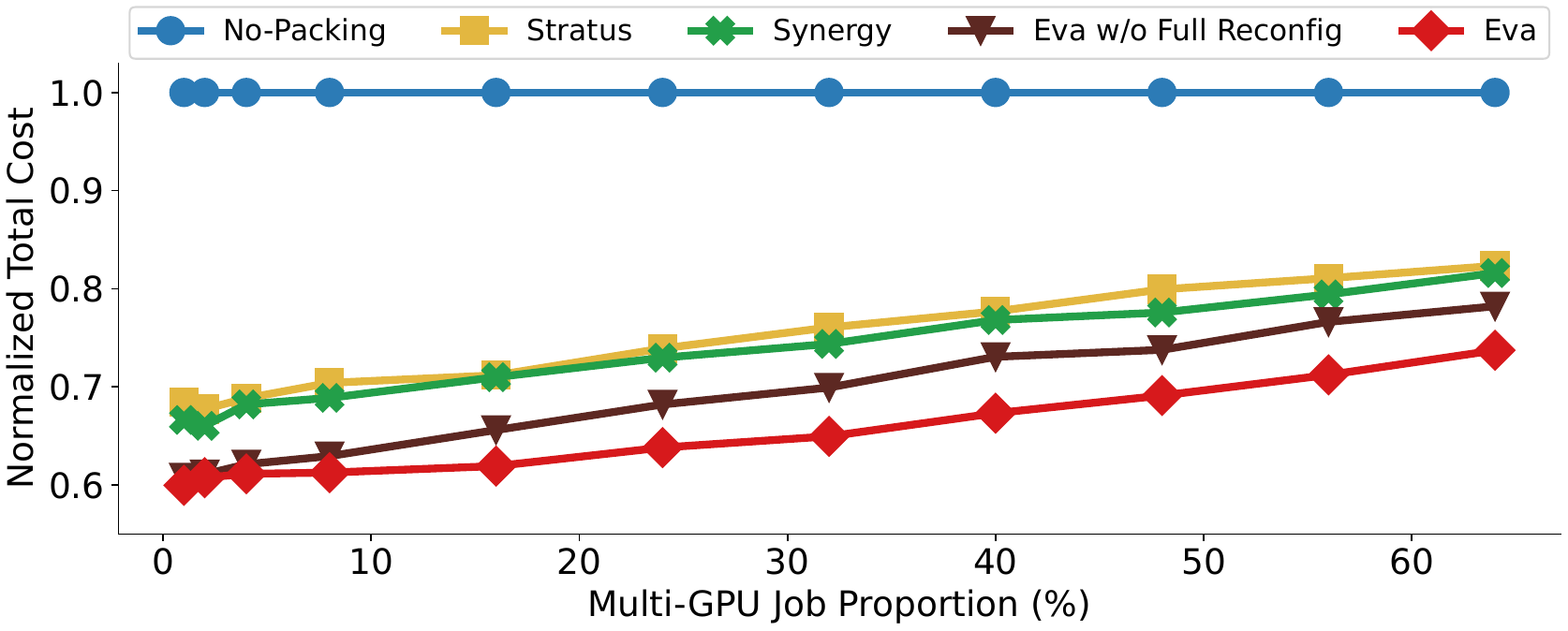}
    \caption{Impact of workload composition.\label{fig:vary_multigpu_jobs/total_cost}}
    \vspace{-1ex}
\end{figure}
As shown in Table~\ref{tab:alibaba-trace-job-composition}, jobs requiring more than a single GPU only accounts for 0.42\% of all jobs in the trace. Single-GPU jobs can be co-located with other jobs on the same instance easily, creating opportunities for cost reduction through packing. We are interested in examining how cost savings are affected when the workload contains a higher proportion of multi-GPU jobs, which offer fewer packing opportunities.

We modify the workload composition to include various proportions of multi-GPU jobs, maintaining a ratio of 5:4:1 for the amount of 2-GPU, 4-GPU, and 8-GPU jobs, which is similar to the relative proportions in the original trace. The proportion of non-GPU jobs remains the same. As shown in Figure~\ref{fig:vary_multigpu_jobs/total_cost}, as the proportion of multi-GPU jobs increases, all packing schedulers experience diminished cost reduction due to the increased difficulty in packing. However, Eva continues to reduce the total cost by 10-15\% compared to Stratus and Synergy.

In \S\ref{subsec:eval/migration}, we see that Full Reconfiguration is adopted less than 12\% of the time. This raises the question of whether Partial Reconfiguration alone could be sufficient. Figure~\ref{fig:vary_multigpu_jobs/total_cost} shows that without Full Reconfiguration, the overall cost could increase by as much as 8\%. The increase in cost is especially significant when there are more multi-GPU jobs in the trace, as achieving the optimal cluster configuration without migrating existing jobs becomes less likely. It is thus important to consider both Full and Partial Reconfiguration in scheduling in order to achieve minimal cost.
\vspace{-2ex}
\subsection{Impact of Multi-task Jobs}
\label{subsec:eval/multi-task}
\begin{figure}[t]
    \centering
    \includegraphics[width=0.93\linewidth]{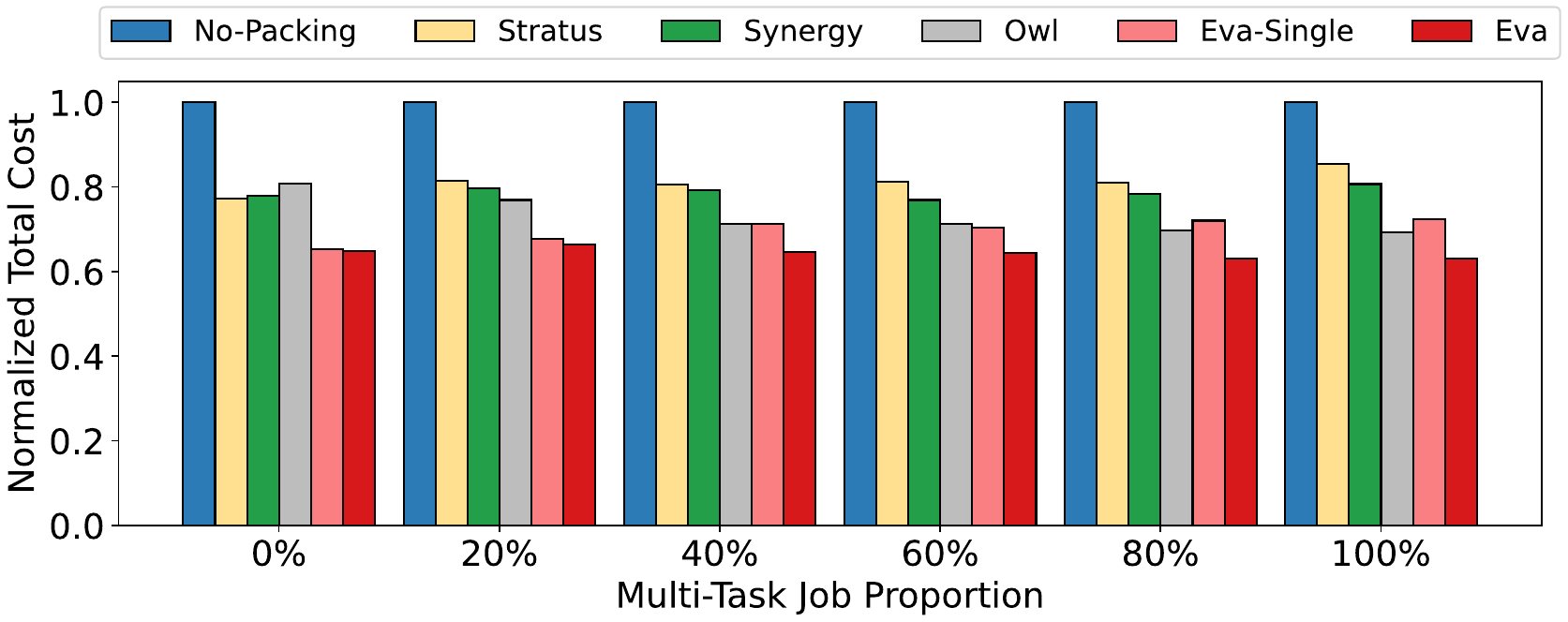}
    \caption{Impact of multi-task jobs.}
    \label{fig:vary_multitask/total_cost}
\end{figure}
As mentioned in \S\ref{subsec:eval/setup}, the Alibaba trace contains only single-task jobs. To introduce multi-task jobs, we modify the trace by randomly selecting a subset of jobs and duplicating their tasks, creating jobs with either 2 or 4 tasks, each maintaining the resource demands of the original task. We vary the proportion of multi-task jobs in the trace while maintaining a 1:1 ratio between 2-task and 4-task jobs. As shown in Figure~\ref{fig:vary_multitask/total_cost}, Eva consistently reduces the total cost by 10-37\% compared to existing schedulers. In addition, we compare Eva with and without considering interdependency within multi-task jobs: \texttt{Eva} and \texttt{Eva-Single}. \texttt{Eva-Single} incur up to 13\% higher costs, reinforcing the results presented in Table~\ref{tab:multi-task-micro-benchmark}.
\vspace{-1ex}
\subsection{Impact of Job Arrival Rate}
\label{subsec:eval/arrival}
\begin{figure}[t]
    \centering
    \includegraphics[width=0.93\linewidth]{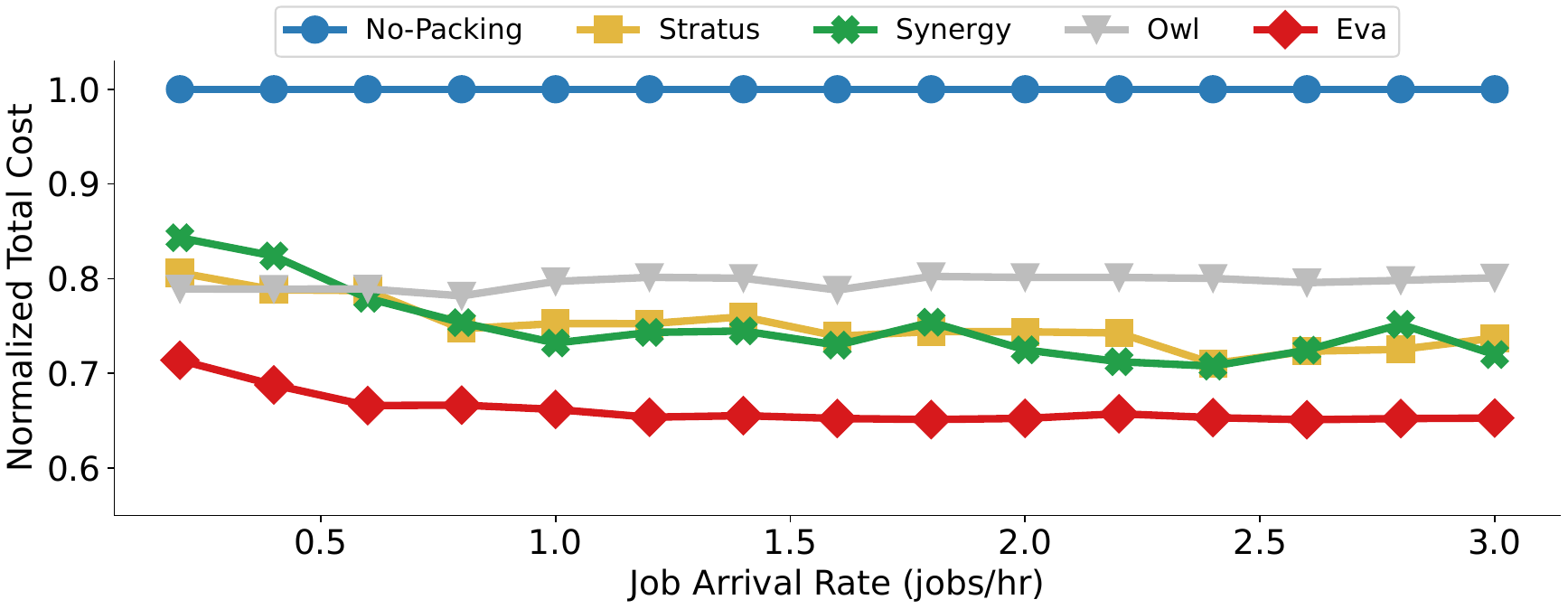}
    \caption{Impact of job arrival rate.\label{fig:vary_job_arrival/total_cost}}
\end{figure}
Figure~\ref{fig:vary_job_arrival/total_cost} shows how job arrival rate affects the benefits of Eva. With a lower job arrival rate, there are fewer jobs in the system at any given time, reducing the opportunities for job-packing. Consequently, packing schedulers achieve less benefits compared to No-Packing Scheduler. However, regardless of the job arrival rate, Eva consistently achieves 10-16\% lower costs than other packing schedulers.
\section{Related Work}
\label{sec:related_work}
\textbf{Cloud-based Cluster Scheduling} Prior work has explored reducing the number of provisioned instances through task co-location but lacks a comprehensive approach for cloud users that effectively accounts for heterogeneous instances~\cite{tchana2015europar, jiang2018self}, migration awareness~\cite{yan2022energy}, and interference awareness~\cite{khan2020consolidation}.

Stratus~\cite{chung2018stratus} addresses the same problem of minimizing total cost in a cloud-based cluster. Designed for interactive and short-running workloads, its scheduling algorithm is conservative in job migration. As discussed in \S\ref{subsec:background/batch_computing}, this gives up potential provision savings when serving long-running jobs. HTAS~\cite{zhong2020htas} builds upon Stratus by segregating interactive jobs from long-running jobs to further reduce mismatch of durations of co-located jobs, but also suffers from conservative migration. In addition, they do not consider co-location interference.

There are cloud-based cluster managers that reduce provisioning cost by taking advantage of cheaper, preemptible spot instances in IaaS cloud~\cite{shastri2017hotspot, harlap2017proteus, xu2019ispot} or price difference between clouds~\cite{yang2023skypilot}. These are orthogonal to our work but could be interesting directions for extensions. However, they do not consider task packing, reducing to the No-Packing Scheduler in our baseline.\\
\textbf{Fixed-sized Cluster} There has been extensive research on cluster scheduling in the context of fixed-sized, multi-resource clusters~\cite{grandl2014tetris, delimitrou2013paragon, ghodsi2011drf, hindman2011mesos, grandl2016graphene}. Recent work on cluster scheduling focuses on serving ML training workload and ensuring high utilization of costly accelerators~\cite{gu2019tiresias, mahajan2020themis, narayanan2020gavel, zheng2023shockwave,qiao2021pollux, mohan2022synergy}. Our work builds upon insights gained from these studies and applies them to cloud-based cluster scheduling which targets minimizing overall cost.\\
\textbf{Co-location Interference} Prior work has attempted to account for the effect of shared resource contention in scheduling. Methods based on low-level hardware counters~\cite{zhuravlev2010contention, nishtala2020hpca} is not applicable in IaaS cloud as these counters are not available on most instance types~\cite{gregg2017pmcs}. Other systems~\cite{mars2011bubbleup, delimitrou2013paragon, delimitrou2014quasar, tian2022owl} predict or directly measure the performance degradation based on profiling. Eva tracks and learns co-location interference online to avoid expensive profiling cost and incorporate this in scheduling to ensure cost-efficiency.\\
\textbf{Dynamic Reconfiguration} 
Dynamic reconfiguration or re-planning is important for environments with fluctuating, unpredictable resources and workloads. This principle extends to various applications, including analytics serving~\cite{ananthanarayanan2014grass},video analytics serving~\cite{romero2021llama}, ML inference serving~\cite{zhang2023shepherd}, and database query optimization~\cite{mahajan2018qoop}. QOOP~\cite{mahajan2018qoop} recalculates query execution plans in response to changes in available resources, switching to a new plan only if the reduction in query execution time justifies foregoing already completed work. Eva follows a similar approach to quantitatively decide whether the decrease in provisioning cost justifies the incurred migration overhead during cluster reconfiguration.
\vspace{-1ex}
\section{Conclusion}
\label{sec:conclusion}
We proposed Eva, a cloud-based cluster scheduler designed to serve batch processing workloads cost-efficiently. Eva employs a reservation price-based scheduling algorithm to jointly optimize task assignment and instance provisioning for minimal cost, and extends the algorithm to incorporate interference awareness and migration awareness. Our physical and simulated experiments show that Eva can reduce costs by 42\% while incurring only a 15\% increase in JCT, compared to provisioning a separate instance for each task.

\section*{Acknowledgements}
We would like to thank our shepherd, John Wilkes, and the anonymous reviewers for their invaluable feedback, which greatly improved our paper. This work was supported by NSF Award CNS-2237306.

\bibliographystyle{plain}
\bibliography{references}

%

\appendix
\section{Artifact Appendix}
\subsection{Abstract}
We have released the artifacts for Eva on Zenodo\footnote{\label{footnote:artifacts_link}\href{https://doi.org/10.5281/zenodo.14880707}{https://doi.org/10.5281/zenodo.14880707}} and GitHub\footnote{\label{footnote:artifacts_link}\href{https://pages.cs.wisc.edu/~tau\_chang/eva}{https://pages.cs.wisc.edu/\textasciitilde tau\_chang/eva}}. In the repository, we provide instructions of setting up Eva on AWS EC2 and S3, along with a minimal working example involving three jobs and four cloud instances to demonstrate Eva's functionality. The simulator and traces used in evaluation are also included.

\subsection{Description \& Requirements}
\subsubsection{How to access}
The source code is available on GitHub. The README provides detailed instructions for setting up Eva on AWS EC2 and S3.

\subsubsection{Hardware dependencies}
The physical experiments in this paper were conducted using AWS EC2 instances, specifically P3, C7i, and R7i. The simulation experiments can be run on any machines.

\subsubsection{Software dependencies}
Artifact software dependencies and the specific versions used for the paper experiments are listed in the GitHub repository README file.

\subsubsection{Benchmarks} 
Table~\ref{tab:evaluate-workload} lists the workload and datasets used in our experiment in Section~\ref{sec:eval}. For the minimial running example (E1), three jobs are hosted on the cloud-based cluster: ResNet18-2 Tasks, GraphSAGE, and A3C. Their execution scripts are included in the repository.

\subsection{Set-up}
The README provides detailed instructions for setting up Eva on AWS EC2 and S3. 
\subsection{Evaluation workflow}
\subsubsection{Major Claims}
\begin{itemize}
    \item (C1): Eva achieves cost saving through task co-location, as shown in Section~\ref{subsec:eval/physical}. This is proven by Experiment 1 (E1).
    \item (C2): Eva reduces the cost of hosting batch jobs on public cloud by 11-42\% compared to existing schedulers, as shown in Section~\ref{subsec/simulation}. This is proven by Experiment 2 (E2) and Experiment 3 (E3).
\end{itemize}

\subsubsection{Experiments}
For each experiment, we provide script to automatically run the experiments. For detailed instruction, please see README.
~\\\\
\textit{Experiment (E1): Small Scale Physical Experiment [20 minutes]:}\\\\
In this experiment, three batch jobs (with a total of four tasks) are submitted and hosted on the cloud-based cluster to demonstrate the functionality of Eva, including task co-location, throughput monitoring and task migration. Experiments can be launched by running \texttt{bash run\_physical.sh} in \texttt{eva/src}.
~\\\\
\textit{Experiment (E2): Comparison with Baselines: Simulation on Partial Alibaba Trace [20 minutes]:}\\\\
In this experiment, we run simulation on the first 200 jobs of the Alibaba trace with all 5 schedulers shown in Section~\ref{subsec:eval/setup}. Experiments can be launched by running \texttt{python experiment\_driver\_200.py} in \texttt{eva/src}.
~\\\\
\textit{Experiment (E3): Comparison with Baselines: Simulation on Full Alibaba Trace [6 hours]:}\\\\
In this experiment, we run simulation on the full Alibaba trace with all 5 schedulers shown in Section~\ref{subsec:eval/setup} to reproduce Table~\ref{tab:simulation-results}. Experiments can be launched by running \texttt{python experiment\_driver\_full.py} in \texttt{eva/src}.

\end{document}